\documentclass[preprint]{rmaa}

\title{Properties of disk galaxies in a hierarchical formation scenario}

\author{Vladimir Avila-Reese \affil{Instituto de
  Astronom\'\i a, Universidad Nacional Aut\'onoma de M\'exico,} 
\and 
Claudio Firmani \affil{Instituto de Astronom\'\i a, Universidad Nacional Aut\'onoma 
de M\'exico,  and Osservatorio Astronomico di Brera}}

\fulladdresses{
\item V. Avila-Reese: Instituto de Astronom\'\i a-UNAM, A.P. 70-264,
04510 M\'exico, D. F., M\'exico (Email: avila@astroscu.unam.mx)
\item C. Firmani: Osservatorio Astronomico di Brera, via E.Bianchi 
46, I-23807 Merate, Italy, and Instituto de Astronom\'\i a-UNAM, A.P. 70-264,
04510 M\'exico, D. F., M\'exico (Email: firmani@astroscu.unam.mx)}

\shortauthor{Avila-Reese \& Firmani}
\shorttitle{Properties of disk galaxies}


\abstract{ We used galaxy evolutionary models in a hierarchical inside-out 
disk formation scenario to study the origin of the main local and 
global properties of disk galaxies as well as some of the 
correlations between their global properties. We found that most 
of disk galaxy properties and their correlations
are the result of three (cosmological) initial factors of the 
system and their dispersions: the virial mass, the halo mass 
aggregation history (MAH), and the angular momentum given through 
the spin parameter $\lambda$. The MAH determines mainly the dark
halo structure and the integral color indexes while $\lambda$ 
determines mainly the disk surface brightness and the 
bulge-to-disk ratio. We calculated star formation using a 
gravitational instability criterion and a self-regulation mechanism 
driven by the energy balance in the disk turbulent interstellar 
medium. The efficiency of star formation in this model is almost 
independent from the mass of the system. Therefore, the galaxy 
integral color index $B-V$ or the stellar mass-to-luminosity ratio,
$M/L_B$, do not depend upon mass. We show that the luminosity-dependent 
dust absorption empirically determined by Wang \& Heckman 
explains the observed color-magnitude and color Tully-Fisher (TF) 
relations without the necessity of introducing a mass-dependent star 
formation efficiency. The star formation history in our models is driven by 
both the MAH and the surface density ($\lambda$). The disks
in centrifugal equilibrium form within growing cold dark matter
halos with a gas accretion rate proportional to the halo MAH. The
disks present exponential surface density and brightness profiles, 
negative radial color index gradients, and nearly flat rotation 
curves. We also calculated the 
secular formation of a bulge due to gravitational instabilities 
in the stellar disk. The intensive properties of our models agree 
very well with the observational data and they follow the same 
trends galaxies follow across the Hubble sequence. The infrared TF 
relations and the luminosity-radius relation predicted for the 
cosmological model used here also agree with the observational 
data. The main shortcomings of the inside-out hierarchical 
formation models presented in this paper are the excessive radial 
color gradients and the dark halo dominion in the rotation curve 
decompositions.}

\resumen{}
\ReceivedDate{October 14, 1999}
\nonstopmode
\begin{document}
\maketitle

\section{Introduction}
The study of galaxy formation and evolution may be approached 
in a deductive or in an inductive way
(e.g., Renzini 1994). In the former case,
starting from some ab-initio conditions given by a theory of cosmic 
structure formation, one tries to model the galaxy evolution in order 
to predict the observable properties of the galaxies. In the latter
case, starting from the present-day properties
of galaxies, and through galaxy evolutionary models, one tries to
reconstruct the initial conditions of galaxy formation. In this paper
we introduce models of disk galaxy formation and evolution based
on both approaches.

{\bf Deductive approach.} Most of current models about cosmic structure 
formation are based on the
gravitational paradigm and the inflationary cold dark matter (CDM)
cosmology. In these models, cosmic structures build up
hierarchically by a continuous mass aggregation process. From the point of
view of galaxy cosmogony, a key question is whether this aggregation
occurs through violent mergers of collapsed substructures or through a
gentle process of mass accretion. According to this, two could be the
most general scenarios of large-galaxy formation. (i) In one case, the 
main properties of luminous galaxies, including those which define 
their morphological types, are
supposed to be basically result of a given sequence of mergers. This
picture ---that we shall call the{\it \ merger scenario}--- has been widely
applied in semianalytical models of galaxy formation, where the global
properties of galaxies are calculated from the cosmological dark 
matter (DM) halo merging
histories using preconceived recipes for the gas cooling, disk mergers,
star formation (SF), supernova (SN) feedback, etc. (e.g., Lacey et al. 1993;
Kauffmann, White, \& Guiderdoni 1993; Cole et al. 1994; Heyl et al. 1995;
Kauffmann 1995; 1996; Baugh, Cole, \& Frenk 1996; Baugh et al. 1997;
Somerville \& Primack 1999). (ii) In the other case, the formation and evolution
of galaxies is related to a gentle and coherent process of mass aggregation
dictated by the shapes of the density profiles of the primordial
fluctuations: galaxies coherently grow inside-out. We shall call this picture 
---firstly developed by Gunn (1981, 1987) and by Ryden \& Gunn (1987)--- the
{\it extended collapse scenario. }Since disk galaxies ($\sim 80\%$ of
present-day normal galaxies) could not have suffered major mergers due to
the dynamical fragility of the disks (T\'{o}th \& Ostriker 1992; Weinberg
1998), the extended collapse scenario appears to be 
more appropriate to study their evolution.

The question of whether the major mergers dominate or not in the assembly
of protogalaxies depends on the statistical
properties of the density fluctuation field. However, even if the DM
structures assemble through chaotic and violent mergers of
substructures, the baryon gas tends to aggregate and collapse in a more
(spatially) uniform fashion than DM do it (e.g., Kepner 1997).
On the other hand, the reionization of the universe probably heated
the gas at high redshifts to temperatures enough to prevent further its
collapse within the small DM substructures of the hierarchy (e.g., 
Blanchard, Valls-Gabaud, \& Mamon 1992; Prunet \& Blanchard 1999). This
way, large galaxies could have formed smoothly with the gas trapped into the 
potential wells of their surrounding DM halos, avoiding an assembly
through the merging of baryon substructures. Only in high density
regions of the universe, as in clusters of galaxies, luminous galaxies 
could have efficiently interacted among each other, thickening and disrupting
their disks and probably forming lenticular and elliptical systems. Disk 
galaxies indeed typically are isolated objects. On the
other hand, the interpretation of high velocity clouds as build-blocks 
of the formation of galaxies in the
Local Group (Blitz et al. 1999) is one possible observational evidence
for the extended collapse scenario: these clouds might constitute
large reservoirs of cold gas which is smoothly accreted onto the 
galaxies (L\'opez-Corredoira, Beckman, \& Casuso 1999).

{\bf Inductive approach.} Galaxy evolutionary models have
shown that due to the relatively rapid disk gas consumption in SF, closed models are
not able to explain several properties of disk galaxies, as well as the wide
range of colors, gas fractions, etc. that galaxies present across the Hubble
sequence (e.g., Larson \& Tinsley 1978; Tinsley 1980; Larson, Tinsley, \&
Caldwell 1980; Kennicutt 1983; Gallagher, Hunter, \& Tutukov 1984; Firmani
\& Tutukov 1992, 1994). On the other hand, it was shown that the SF time
scale in disk galaxies is not mainly controlled by the initial gas surface density
(Kennicutt 1983; Firmani \& Tutukov 1994; Kennicutt, Tamblyn, \& Congdon
1994). Hence, models where accretion of fresh gas is introduced are 
more realistic. What is more, in order to reproduce the 
radial abundance gradient of the Galaxy, the gas 
accretion should be such as an inside-out disk formation scenario
(e.g., Portinari \& Chiosi 1999, and more references therein). Gas accretion 
might also be necessary to maintain spiral structures. In the
case of open (infall) models, galaxy formation and evolution could
be two related processes where the SF time scale is mainly driven by the 
gas accretion rate at which the disk is being built up.

{\bf Gas infall and SF.} The gas infall rate onto the disks may be controlled 
by the global process of galaxy formation (cosmological accretion) and/or by a
self-regulated process of SF. This latter process, proposed
in the seminal papers by White \& Rees (1978) and White \& Frenk (1991), 
is commonly applied in the merger scenario models. According to 
this mechanism, the gas accretion rate is driven by the cooling 
of the hot gas corona sustained by
the SN-injected energy. However, the self-regulated disk-halo SF models
suffer from some inconsistencies. As Nulsen \& Fabian (1996) pointed out, SN
feedback over large scales occurs on roughly the same time scales as the SF,
not fast enough to tightly regulate the SF rate. On the other hand, because 
the disk turbulent interstellar medium is a very dissipative system (Avila-Reese \& 
V\'azquez-Semadeni 2000), gas and energy outflows tend to be confined 
into a region close to the disk. The thick gaseous disk and the global magnetic 
fields are efficient shields that prevent any feedback toward the halo on large 
scales (e.g., Mac Low,  McCray, \& Norman 1989; Slavin \& Cox 1992;
Franco et al. 1995). The lack of observational evidence for significant 
amounts of hot gas in the halos of disk galaxies confirms this 
picture and suggests that the halo-disk connection is not enough to 
self-regulate and drive the disk SF at the halo scales. 

{\bf An unified scenario of galaxy formation.} The galaxy infall models 
suggested by the inductive approach can in a natural fashion be 
incorporated into the cosmological (deductive) extended
collapse scenario. In Avila-Reese, Firmani, \& Hern\'{a}ndez (1998,
hereafter AFH98), starting from the cosmological settings, the formation and
evolution of galaxy DM halos were calculated through a simplified 
approach whose results are in good agreement with those of the
cosmological N-body simulations (see also Avila-Reese et al. 1999). 
In Firmani \& Avila-Reese (2000a, hereafter FA), the formation and
evolution of baryon disks within the growing halos produced with 
this approach was calculated
in order to study the structural and dynamical properties of disk galaxies. 
Here, we apply a similar method including bulge formation, with the aim
to explore whether the origin of the main properties of disk galaxies and 
their correlations (the Hubble sequence, for example) can be related or not 
to cosmological initial conditions and factors.  In an acompanying paper 
(Firmani \& Avila-Reese 2000b) results about the evolution of the model 
galaxies will be presented and compared with the observational data.

In the extended collapse scenario we develop here, the relevant 
assumptions we introduce are:  (1) spherical symmetry and
adiabatic invariance in the gravitational collapse of
DM halos, (2) spin parameter $\lambda$ produced by cosmological torques
and constant in time for a given halo, (3) gentle incorporation 
of the gas (no mergers) into a disk in centrifugal equilibrium, 
(4) detailed angular momentum conservation and adiabatic invariance
during the gas collapse, (5) stationary SF induced by global disk
gravitational instabilities and regulated by an energy balance {\it within the
disk} interstellar medium, and (6) formation of a bulge due to gravitational 
instabilities in the stellar disk (secular formation scenario).  

A consequence of item (5) is that in our models the SF efficiency is almost 
independent from the galaxy mass or circular velocity. An important   
question is whether the observed ``color'' Tully-Fisher or 
color-magnitude relations can be explained without invoking
SF efficiency and/or disk-to-total mass ratios depending upon mass 
(luminosity). We propose that the observed luminosity-dependent 
dust absorption (e.g., Wang \& Heckman 1996) explain these 
empirical relations.

In the last years, several methods to model galaxy formation and evolution
were proposed: the numerical simulations (N-body+hydrodynamics),
the semianalytical method, and the analytical method. We have called 
our method {\it seminumerical}, essentially with the aim to differentiate it 
from the other ones. In fact, the virialization of the halos is
calculated with an itterative procedure and the disk formation and
evolution is followed by solving the hydrodynamical equations (in the
radial direction) with a full numerical method. As in the 
numerical  simulations, with the 
seminumerical method, we try to model the evolution of the {\it internal} 
structural, dynamical, kinematical and SF processes of an individual
galaxy (dark+luminous). On the other hand, we are also interested in 
calculating statistical properties of a whole population of galaxies. 
The techniques used in semianalytical and analytical methods have 
proved to be especially powerful in this undertaking. In our method 
we use some related techniques. It is important to mention that, 
from the physical point of view, our seminumerical 
method and the analytical method are similar: both are based on 
the extended collapse 
scenario.  The results obtained with the analytical method in recent 
papers (Dalcanton, Spergel, \& Summers 1997; Mo, Mao, \& White 1998; 
van den Bosch 1998, 1999) encourage us to study in more detail this scenario,  
following the overall process of formation and evolution of individual
disk galaxies (dark+luminous).

The plan of the paper is as follows: in $\S$ 2 we describe our method 
and explain the strategy in order to calculate the galaxy models; in 
$\S \S$ 3 and 4 we present the local and global properties of the model 
galaxies at $z=0$, respectively, and compare them with the observational data; 
in $\S$ 5 we study some important correlations among the global
properties of the models, particularly those which go across the Huble
sequence, and we explore whether the ``color'' Tully-Fisher
and color-magnitude relations can be explained by the 
luminosity-dependent dust absortpion;
the fundamental factors of the galaxy models and the way they determine
the galaxy properties are presented in $\S$ 6; we discuss the main difficulties 
found in the extended collapse scenario in $\S$ 7, and  
finally, in $\S$ 8, we summarize the paper and present the conclusions.

\section{The method}

Following, we briefly describe the {\it seminumerical}  method 
used to calculate the 
formation and evolution of isolated disk galaxies in the extended collapse 
scenario based on the hierarchical formation picture. Most of the steps of 
this method were already described in previous works (Firmani et al. 
1996; Avila-Reese 1998; AFH98; FA). 

{\bf Virialization of the DM halos.} The first step in our model is 
to calculate 
the gravitational collapse and virialization of isolated DM halos starting 
from a primordial density fluctuation field. This was done in AFH98 and in
Avila-Reese 1998. In order to generate the hierarchical MAHs of the DM halos
we use the extended Press-Schechter approximation based
on the conditional probabilities for a Gaussian random  field (Bower 1991; Bond 
et al. 1991; Lacey \& Cole 1993). For a given present-day mass $M_{\rm nom}$, 
we generate a set of MAHs through Monte Carlo simulations. We 
follow the aggregation history of the main progenitor by identifying 
the most massive subunit of the distribution at each time. The 
gravitational collapse and virialization of the DM halos is
calculated with an iterative seminumerical method, assuming spherical 
symmetry and adiabatic invariance during the collapse. This method is 
based on the secondary infall model (e.g., Gunn 1977; Zaroubi \&
Hoffman 1993), and it expands upon it by allowing non-radial motions 
and arbitrary initial conditions (MAHs in our case). 

Not all the mass $M_{\rm nom}$ virializes at $z=0$. The mass shells 
that are already virialized roughly correspond to those within the 
virial radius $r_v$ at which the 
halo mean overdensity drops below the critical value $\Delta _c$
given by the spherical collapse model. For the cosmological model used here 
$\Delta _c=179$ (e.g., Padmanabhan 1993). Analysis of halos identified in 
numerical simulations confirm this result (Cole \& Lacey 1996; 
Eke, Navarro, \& Frenk 1998). The mass contained
within $r_v$ is the virial mass $M_v$ which, depending upon the MAH, 
is equal to 0.7-0.9 times $M_{\rm nom}$ (see also Kull 1999).

The halos calculated with our method present a diversity of density profiles,
the most typical being close to that suggested by Navarro, Frenk, \& White
(1996, 1997) (AFH98).  This diversity
in the structure of the halos is related to the dispersion in the MAHs.
As it was tested for a flat CDM model with cosmological constant, the diverse
density profiles obtained with our method agree rather well with the results 
of cosmological N-body simulations (Avila-Reese et al. 1999).  
 
{\bf Disk build-up.} 
Baryon matter is able to cool and further collapse within 
the virialized DM halo. The collapse can be halted only by
centrifugal forces or when gas is transformed into
stars. While the second case could be related to the origin of large bulges
and elliptical galaxies, the former is commonly invoked to explain the
formation of disk galaxies. The build-up of disks within the evolving DM
halos is carried out as follows: 

(1) We consider that, at the beginning,
baryon matter has mass and angular momentum distributions similar to those
of the DM; it is assumed that the spherical shells are in solid body
rotation. 

(2) Once the current mass shell has attained its maximum expansion
radius, a fraction $f_d$ of its mass is transferred in a virialization time
into a disk in centrifugal equilibrium. This mass fraction is assumed to be 
{\it only in form of gas}. Since for galaxy halos the time scale
of gas cooling is generally smaller than the dynamical time scale, the 
shell virialization time is a reasonable ---probably minimum--- time 
scale for gas incorporation into the disk, considering that the gas 
will be shock-heated during the virialization process. The radial 
mass distribution of the disk gas is
calculated equating its specific angular momentum to the one of 
its final circular orbit (detailed angular momentum conservation 
is assumed). The specific angular momentum  $j_{sh}$ acquired by
each collapsing shell during the linear regime is estimated under the
assumption of a constant spin parameter $\lambda $ 
($\equiv  (J\left| E\right| ^{1/2})/(GM^{5/2})$): 
\begin{equation}  
j_{sh}(t)=\frac{dJ(t)}{dM(t)}=\lambda \frac{{GM(t)^{5/2}}}{\left|
E(t)\right| ^{1/2}}\Bigl(\frac 52\frac 1{M(t)}+
\frac{d\left| E(t)\right| }{2dM(t)}\Bigr)
\end{equation}
where $M$, $J$ and $E$ are the total mass, angular momentum and energy of the
object at time $t$, respectively. According to analytical studies and to the 
outcomes of cosmological N-body simulations, the DM halos have values 
of $\lambda$ given by a log normal distribution with an
average of $\sim $0.05 and a width in the logarithm less than one 
(e.g., Catelan \& Theuns 1996 and references therein).

(3) The gravitational drag on the total system 
produced by the central disk is calculated at each time with 
the adiabatic invariant formalism (e.g., Flores et al. 1993).

{\bf Disk star formation and hydrodynamics.} In our model, the 
SF and the internal hydrodynamics of the disk are
regulated by an energy balance between the SN and gas accretion
kinetic energy input, and the turbulent energy gas dissipation (Firmani et
al. 1996). Star formation is turned on at radius $r$ when the 
Toomre gravitational instability parameter for the gas disk,
\begin{equation} 
Q_g(r)\equiv \frac{v_g(r)\kappa (r) }{\pi G\Sigma _g(r)},
\end{equation}
falls below a given threshold (Toomre 1964); $\kappa $ is the epicyclic 
frequency, $v_g$ the gas rms turbulent velocity, and $\Sigma _g$ the gas 
surface density. Thus, the SF is controlled by a feedback mechanism such
that, when a gas disk column is overheated by the SF activity, SF is inhibited and
the disk column dissipates the excess energy to lower $v_g$ back to the value
determined from the Toomre criterion threshold. Owing to several theoretical
and observational pieces of evidence (see references in FA)  we fix
$Q_g=2$ instead of the value of 1 given by Toomre (1964) for an infinitely
thin disk. While the SF rate is rather insensitive to $Q_g$, the disk thickness
strongly depends upon it. When $Q_g=2$ is used, for a model of the Galaxy
we obtain the observed gas and stellar thicknesses. The gas loss from 
stars is also included. The gravitational dynamics 
of the evolving star and gas disks, and the DM halo are treated in detail.  

The local disk galaxy models presented in Firmani \& Tutukov (1994) included
the integration of luminosities in $B$ and $V$ bands for all the 
stellar population.
A Salpeter initial mass function with a minimal star mass of 0.1M$_{\odot }$, 
and solar metallicities were used. Here, the results from these models 
are used to calculate the surface $B-$ and $V-$band brightness  at 
every radius and throughout the evolution of the galaxy models (see Firmani 
et al. 1996). We find that the approximation to the population synthesis we
use provides $B-V$ colors that can be less by $\sim 0.1$ mag in the red,
and by $\sim 0.05$ 
mag in the blue than the respective values given by more sophisticated
models (e.g., Bruzual \& Charlot 1993; Charlot, Worthey, \& Bressan 1996).

According to our scheme, time by time 
and at each radius, the growing disk is characterized 
by the infall rate of fresh gas by unit of area, 
$\dot {\Sigma _g}(r,t)$, the gas and stellar disk surface 
density profiles, ${\Sigma _g}(r,t)$ and ${\Sigma _s}(r,t)$, 
the total rotation curve (including the growing DM halo 
component), $V_r(r,t)$, and the SF rate $\dot {\Sigma} _s(r,t)$ 
determined by the energy balance in the vertical 
gaseous disk and by a Toomre criterion.
It is important to note that in our self-regulating SF mechanism the feedback
happens only {\it within} the disk and not at the level of the whole halo.
This is justifiable, as was already explained in the Introduction, because
the galaxy disks are dense and very dissipative systems that confine
gas and energy outflows.

{\bf Bulge formation.} Recent observational and theoretical studies are 
challenging common preconceptions about galactic bulges; for a review 
see Wyse, Gilmore, \& Franx (1997). These studies tend to show that 
several bulge formation
mechanisms could be working in galaxies. Bulges in galaxies with low and
intermediate bulge-to-disk (b/d) ratios may be formed through secular
dynamical evolution of the disks, whereas bulges in galaxies with large b/d
ratios could have been formed separately from the disks, through an early
dissipative collapse, and/or from mergers.

In order to estimate the bulge mass we
have introduced a recipe according to which the stars of the central
``cold'' disk region where the stellar Toomre instability parameter $%
Q_s(r)\equiv \frac{v_s(r)\kappa (r)}{3.36G\Sigma _s(r)}$ is less 
than 1, are transferred to a spherical component in such
a way that $Q_s$ remains equal to 1; $v_s$ and $\Sigma _s$ are 
the radial stellar velocity dispersion and stellar surface density. The
physical sense of this recipe is in agreement with the {\it secular} scenario of
bulge formation where gravitational instabilities in the stellar disks
produce bars which dissolve forming a ``hot'' component, the bulge (see
Norman, Sellwood, \& Hasan 1996, and references therein). Indeed, the
similarity found in colors (Peletier \& Balcells 1996), and scalelengths (de
Jong 1996a; Courteau, de Jong, \& Broeils 1997) between disk and bulge 
for late-type galaxies might
mean that the disk and the bulge are closely related. It is worth emphasizing
that our recipe of bulge formation is a very crude approximation to a
complex phenomenon about which, in fact, not much is known, both from the
theoretical and observational points of view.

\subsection{Selection of the models}

The {\it seminumerical} method described above allows us to calculate 
the properties of disk galaxies at any time, particularly at the present-day, 
starting from the cosmological initial conditions. Since our aim is to study 
general behaviors, here we shall use only one representative cosmology,
the standard CDM model with the following parameters: matter density 
$\Omega _m=1$, baryon fraction density $\Omega _b=0.05$, Hubble parameter 
$H_0=h\times 100$ kms$^{-1}$Mpc$^{-1}$ with $h=0.5$, the amplitude 
of fluctuations on $8h^{-1}$Mpc scale $\sigma _8=0.6$. At galaxy 
scales, the power spectrum of this model is similar to that 
of the popular flat CDM model with cosmological constant 
($\Omega _{\Lambda}\approx 0.7$, $h\approx 0.7$) normalized to the $COBE-$
satellite measurements. We calculate the power spectrum according to
Sugiyama (1996). Once the cosmological
model is fixed, the galaxy model to be studied should be defined. 
A given model is characterized  by its (dark+baryon) virialized 
mass $M_v$\footnote{As it was 
explained above, we give as initial parameter the nominal mass 
$M_{\rm nom}$. Nevertheless, not all this mass is virialized; the most 
outer mass shells that encompase this mass are still in the 
process of virialization. We use the virial mass $M_v$ as 
the current  mass of the halo. This mass  is approximately  
$0.7-0.9 M_{\rm nom}$, depending on the MAH.}, its MAH,  its 
initial spin parameter $\lambda $, and by the mass fraction 
$f_d$ of $M_v$ that is incorporated into the disk. In order 
to obtain a complete galaxy population, one should calculate 
models with masses, MAHs, and $\lambda ^{\prime }s$ drawn from 
the corresponding statistical distributions given by the 
cosmological framework. A complete modeling of the disk galaxy 
population, however, is beyond the scope of this paper. Regarding $f_d$, 
it is probably the same for all the galaxies, although 
some astrophysical process could change their initial values (e.g., 
Natarajan 1999). Here we shall assume $f_d$ constant and 
equal to 0.05 (see Mo et al. 1998 and FA for some justifications for taking
this value). This value is within the range predicted by the theory
of light elements production, $\Omega _bh_{0.5}^2 \approx 0.025-0.053$, as
well as by the range given by the global budget of baryons,
$\Omega _bh_{0.5}^2 \approx 0.014-0.080$ (Fukugita, Hogan, \& Peebles 1998).
Note that, taking $f_d=0.05$ for the standard CDM model, we are assuming
that in the galaxy halos almost all the baryon matter is incorporated 
into the disk. 

The aim of this paper is to explore to what extent the extended collapse 
scenario is able to provide the correct initial conditions for galaxy 
evolutionary models. With this aim in mind, we 
shall study and compare with observations (i) the main structural and
luminosity characteristics of galaxies corresponding to a local disk galaxy
population, and (ii) the main correlations among  the global properties,
particularly those which go across the Hubble sequence. To achieve the former item, it 
is enough to calculate models for a significant range of masses, MAHs 
and $\lambda ^{\prime }s$, and test whether the obtained galaxy properties are 
realistic for these ranges. Regarding the latter item, as was mentioned above, 
we shall not model a complete galaxy population; only general 
trends in the correlations among the global galaxy properties will 
be obtained. As a matter of fact, the available observational 
galaxy samples are not complete enough so as to provide complete 
statistical information on galaxy properties and their correlations. 
Only certain relations, as the Tully-Fisher relation (TFR), were 
observationally studied with enough detail as to have a statistical 
description.

We calculate models for three representative masses: $M_{\rm nom}=$
5$\times 10^{10}$M$_{\odot },$ 5$\times 10^{11}$M$_{\odot },$ 
and 5$\times 10^{12}$M$_{\odot }^{3}$. For each mass, three MAHs
are selected from its distribution: the average, and two symmetrical deviations 
such that roughly $80\% $ of all the MAHs are contained between them 
(see AFH98). For each mass and for each MAH, we calculate models taking 
three values of $\lambda :$ 0.035, 0.05, and 0.1. Thus, we obtain 
a grid of 27 models. 

\section{The local properties of the models}

{\bf Local star formation rate.} We obtain that 
\begin{equation} 
\Sigma _{\rm SFR}(r) \propto\Sigma _g^n(r)
\end{equation}
with $n\approx 2$, for most 
of the models and along a major portion of the disks, 
where $\Sigma _{\rm SFR} $ is the SF rate
per unit of area (see also Firmani et al. 1996). This SF law is 
basically a consequence of the self-regulation feedback mechanism 
used in our models (Firmani \& Avila-Reese, in preparation). Other 
self-regulation SF models also yield SF rates with (near to 2) power-law
dependences on local $\Sigma _g$ (see references in Struck 
\& Smith 1999).  Unfortunately, the observational data on the local
SF rate of disks galaxies are still very poor. The data presented in 
Kennicutt (1989, 1998) show that the disks at intermedium radii
typically have a $n\approx 2$ SF law, while at inner radii the
index is typically smaller than 2 and near to 1. 
Our SF prescription in inside-out forming disks is just a simple 
approximation to very complex phenomena; at this level, we think that 
our model predictions are succesful.

{\bf Surface brightness profiles.} The mechanism of disk 
formation used here naturally leads to an exponential radial distribution 
of the disk surface density. Note that we have not assumed 
initial angular momentum distribution corresponding to an object in 
solid body rotation (e.g., Fall \& Efstathiou 1980; 
Dalcanton et al. 1997). The crucial point is that we have assumed 
$\lambda$ constant in time (see eq. 1). Analysis of cosmological 
N-body simulations seem to confirm this assumption (Gottlober, 
personal communication). In Figure 1a we show the surface brightness 
(SB) profiles in the $B$ band for different models of
$M_{\rm nom}=$5$\times 10^{11}$M$_{\odot }$. The SB profiles are nearly 
exponential and the models with large $\lambda$ form low SB galaxies (see also 
Dalcanton et al. 1997; Jim\'enez et al. 1998). For other masses the situation 
is the same, although with the mass the SB increases on average. 

{\bf Radial distribution of the $B-V$ color index.} A natural consequence 
of the inside-out disk galaxy formation mechanism is the existence of 
negative radial color index gradients along the disk. In Figure 1b we plot 
the radial $B-V$ color  distributions for the same models of Figure 1a.
Multiwavelength surface photometry studies of
galaxies (e.g., de Jong 1995,1996b) confirm that galaxy disks tend to be bluer
at the periphery, although these gradients are typically smaller than those
predicted by our models. We shall return to this point in $\S 6.2$.


{\bf Rotation curves.}
In Figure 2, the rotation curves for the same models of Figure 1 are
depicted. The shape of the
rotation curve depends mainly upon the spin parameter $\lambda $ and the 
MAH, and it correlates with the central SB. Observational studies
seem to confirm the fact that galaxies with higher SB have 
more peaked rotation curves (e.g., Casertano \& van Gorkom 1991; 
Tully \& Verheijen 1997;  Verheijen 1997). Models  with the 
early active MAH have more concentrated rotation curves than models 
with an extended MAH. In Figure 2, the rotation curves of the 
$M_{\rm nom}=5\times 10^{10}M_{\odot }$ and 5$\times 10^{12}$M$_{\odot }$ 
models, corresponding to the
average MAHs and $\lambda $, are also depicted. At the Holmberg radius the
less massive models have nearly flat rotation curves, while the more massive
galaxies present decreasing rotation curves. A similar trend was found
from a large observational sample (Persic, Salucci, \& Stel 1996). This 
could seem contradictory with the result that less massive DM halos 
are more concentrated than the more massive ones (Navarro et al. 1996,1997; 
AFH98; Avila-Reese et al. 1999). However, in our models, this is compensated by 
the fact that the {\it disks} in massive halos are more concentrated than in small 
halos (see also Dalcanton et al. 1997), with the result that their 
contribution to the total rotation curve is more significant in 
the more massive halos than in the less massive ones. The disk 
mass fraction $f_d$ also influences on the rotation curve shape. If 
we fix $f_d<0.05$ then the rotation curves of all the model result 
less peaked than those presented in Figure 2 (see also Mo et al. 1998, FA). 


Regarding the decomposition of the rotation curves, as it was already shown in
AFH98 and FA, we obtain that the halo component dominates almost everywhere
for most of the models. This result is in potential disagreement 
with the rotation
curve decompositions inferred from observations (e.g., Carignan \& Freeman 
1985; Sancisi \& van Albada 1987; Begeman 1987; Verheijen 1997; 
Corsini et al. 1998). In $\S$6.3 we shall return to this question.

\section{The global properties of the models}

Among the global properties that our models predict for a disk galaxy, we
shall take into account for this study the integral $B-V$ color 
index (calculated within a Holmberg
radius), the $B$-band luminosity $L_B,$ the $B$-band exponential disk scale
length $h_B$, the $B$-band central SB $\mu _{B_o}$ or $\Sigma
_{B_o}$ ($\mu _{B_o}$ is given in magnitudes per arcsec$^2,$ and $\Sigma
_{B_o}$ in $L_{B_{\odot }}$ per pc$^{-2}$), the disk gas fraction $f_g$ ($%
\equiv \frac{M_{gas}}{M_{gas}+M_{stars}}$), the stellar bulge-to-disk ratio
b/d, and the maximum rotation velocity $V_m.$ The correlation
matrix from a principal component analysis of these properties for the 27
models calculated here is given in Table 1. The correlation coefficients 
{\it give only a qualitative estimate} of the correlations; note that 
the initial parameters of the models were not drawn randomly from the 
corresponding statistical distributions ($\S 2.1)$.


From Table 1 is seen that the observational trends across the Hubble sequence 
are reproduced 
for the intensive properties: the redder and more concentrated is 
the disk, the smaller is the gas fraction and the larger is 
the b/d ratio. Moreover, the models seem to populate a planar
region in the $\mu _{B_o}-(B-V)-f_g$ or $\mu _{B_o}-(B-V)-b/d$ 
spaces, in agreement with observations (McGaugh \& de Blok 1997).
$B-V$ and $\mu _{B_o}$ are almost independent one from the other, 
so that we can express $f_g$ and b/d as functions of these 
two parameters. In Figure 3, $f_g $ is plotted versus $B-V$ 
and $\mu _{B_o}$. The gas fraction is larger for
bluer colors $B-V$ (panel a), which means that $f_g$ is larger for 
the MAHs whose present-day gas infall rate is still high. On the 
other hand, in panel b it is seen how a less concentrated disk 
(larger $\lambda $) presents a higher $f_g$ than a disk with high 
$\Sigma _{B_o}$ (low $\lambda$). On the basis of
this result is the influence of the disk gravitational compression upon the
capability of gas to form stars (Firmani \& Tutukov 1992, 1994). The disk
surface density strongly influences the b/d ratio. This is because the stellar
surface density enters in the Toomre gravitational instability criterion
which is used in our models to calculate the formation of bulges. The larger
the central SB (stellar density), the larger is the b/d
ratio (Fig. 4b). The mass and MAH introduce a dispersion in this
correlation because they influence the other quantities which appear in the
Toomre criterion. In Figure 4a is seen how marginally does b/d depend upon
$B-V$. The observational data show that b/d correlates better with $%
\mu _{B_o}$ than with $B-V$ (e.g., de Jong 1996a).


In Figures 3 and 4 are also plotted the observational data taken from a
compilation presented in McGaugh \& de Block (1997) where low SB 
galaxies are included. The $B-V$ color indexes were not corrected for the internal
(inclination) galaxy extinction. We have applied this correction according
to the formula given in the RC3 catalog (de Vaucouleurs et al. 1991). The
b/d ratios for the  low SB galaxies presented in McGaugh \& de Block 
(1997) were not estimated, so
that in the panels where this ratio is plotted the  low SB galaxies are not
considered. In general, the model properties fall rather well within the 
observational ranges and are in agreement with the empirical correlations 
(compare also Table 1 with the correlation matrix presented in 
McGaugh \& de Block 1997). It is encouraging
that the b/d ratios predicted by the models using the simple gravitational
instability criterion (secular bulge formation) are in agreement with those
inferred from observations (de Jong 1996a,c). Note that de Jong (1996c) used
an exponential profile in his two dimensional bulge-to-disk decomposition
procedure instead of the de Vaucouleour's profile for the bulges, arguing
that such a profile fits better the observations (see also 
Andrekakis \& Sanders 1994; Andrekakis, Peletier, \& Balcells 
1995). That is why the b/d ratios obtained by him are smaller than 
those given by previous b/d decompositions (e.g., Simien \& de 
Vaucouleurs 1986). The bulges of galaxies with b/d ratios larger
than those our models predict can be the product of the collapse
and/or merger mechanisms of bulge formation which we do not treat.



The most serious inconsistency between theory and
observation in Figures 3 and 4 is probably that the color index of the models 
are bluer than those of the observed galaxies. Part of this inconsistency might be
due to the imprecision of our population synthesis model (see $\S 2)$.
Nevertheless, other explanations are also possible. The statistical range ($\sim
80\%)$ of the MAHs calculated here for the Gaussian fluctuations leads to
disks with $B-V$ between approximately $0.4$ and $0.7$. Models 
with colors redder than 0.7 mag may be produced, although only 
for extreme cases. We have found that the
color index becomes very sensitive to the MAH when this corresponds to early
active MAHs: for some extreme cases the $B-V$ color may be as red
as $\sim 0.95$ mag. Thus, some models can easily attain colors redder than
0.7 mag; of course the frequency of such models will be low. In Figures 3 and 4,
with dashed lines we show the range of values of the different properties of the 5$%
\times 10^{11}M_{\odot }$ models for the three $\lambda ^{\prime }s,$ when
the statistical range in the MAHs is symmetrically extended to $94\%$ (symbols
consider only $80\%$). As is seen, the models can be as red as some observed
galaxies are. 

The addition of some extra phenomena in our models 
may help to obtain galaxies redder than those presented in Figures 3 and 4.
For example, the internal extinction in combination with the metallicity-luminosity
relation can introduce an important effect of reddening, particularly for
the most massive galaxies (see \S 5.2). Using the results presented in Wang
\& Heckman (1996) (hereafter WH96), and the Galactic extinction curve for $R_V=3.1$
(Cardelli, Clayton, \& Mathis 1989), we have reddened the models
corresponding to the average MAH and $\lambda =0.05$, represented in Figures
3 and 4 with black filled circles. The error bars account for the range of
parameter values given in WH96 (see Appendix). 

The influence of environment on the galaxy evolution might also 
help to produce models redder than 0.7 mag. In dense environments, 
the gentle mass aggregation is probably truncated early.
Experiments show that if gas accretion is truncated in the models at 6 Gyrs
and 4 Gyrs, then $B-V$ roughly increases by 0.08  and 0.15 mag, 
respectively. On the other hand,
the early interactions in the dense environments can induce non stationary SF
which produces a fast gas consumption into stars. In a recent work, M\'arquez
\& Moles (1999) studied a sample of very isolated galaxies and they found that
the corrected total $B-V$ color index of the galaxies lies between 0.35 and 
0.85 magnitudes with an average value of 0.48 magnitudes, i.e. the very isolated 
galaxies indeed tend to be blue. 

The observational sample presented in Figures 3 and 4 is
statistically incomplete and plagued of observational uncertainties. This
and possible effects of extinction could account for the large
scatters in the empirical correlations among the intensive properties 
with respect to those of the models. Nevertheless, it is probable that
the intrinsic scatters are, in any case, larger than those found in our
simulations. This is because the scenario proposed here takes into 
account only the most relevant ingredients
of galaxy formation and evolution, omiting certain phenomena
which are not dominant, but are observed in some galaxies. For example, the SF
prescription used in the models is basically a stationary process, while in
real galaxies SF sometimes may appears in bursting modes. This fact
introduces a stochastic component in the photometric features, particularly
for low mass galaxies (Firmani \& Tutukov 1994).

\section{Correlations between global properties}

\subsection{The Tully-Fisher relation in the infrared bands and its scatter}

The galaxy luminosity in the infrared bands is a reliable tracer of the disk
stellar mass (e.g., Pierce \& Tully 1992; Gavazzi 1993; Gavazzi, Pierini, \& 
Boselli 1996). Thus, the TFR in the infrared bands may be
translated to a relation between the disk stellar mass $M_s$ and 
the maximum rotation velocity $V_m$ of disk galaxies. In AFH98 
it was already shown that this relation for the  
$\sigma _8\approx 0.6$ SCDM model is in good agreement
with that derived from most of observations. AFH98 assumed that
the total disk mass is directly proportional to $M_s$. Since here 
we calculate the SF, this assumption is relaxed. The $M_s-V_m$ 
relation we obtain is actually very similar to that estimated in 
AFH98. However, the scatter in the relation now is smaller than 
in AFH98. This is because $M_s$ depends upon disk surface density. 
The efficiency of SF is larger in disks with larger 
surface densities. Thus, for a fixed disk total mass (stars+gas), the model 
galaxies with higher surface densities have larger $V_m$ than models
with low surface densities, {\it but} they also have larger 
stellar masses $M_s$. In the $M_s-V_m$ diagram, the models in 
the high velocity side shift to high values of $M_s$, while the
 models in the low velocity side shift to low values of $M_s$. This 
compensating effect is responsible for a scatter in the $M_s-V_m$ 
relation smaller than that we would expect if one assumes that $M_s$ 
is directly proportional to the total disk mass. This effect also 
produces that high and low surface density (brightness) models 
have approximately the same $M_s-V_m$ relation, i.e. the TFR of low 
and high  models is nearly the same (see also FA). 
The observational data point out to the same fact (c.f., Zwaan et al. 
1995; Verheijen 1997).

In our models the rms scatter in the $M_s-V_m$ relation (or TFR) 
is originated by the scatter in the structure of the DM halos 
(due to the scatter in the MAHs), and by the dispersion of 
the spin parameter $\lambda$. The contribution of the latter 
to the total scatter is diminished due to the effect just  mentioned 
above. To estimate the scatter in the $M_s-V_m$ relation we should 
generate a catalog of models where, at least, the MAH and 
$\lambda$ are drawn through Monte Carlo simulations from their 
corresponding statistical distributions. This was done in FA
for a low density CDM model with cosmological constant. At galaxy 
scales the power spectrum of fluctuations of this model is similar to the SCDM
model used here. In FA it was obtained that the quadratic contributions 
to the scatter in the 
$M_s-V_m$ relation due to the scatters in the MAHs and in $\lambda$ are
approximately $75\%$ and $25\%$, respectively. Thus, the scatter in the MAH
is the main source of the scatter in the $M_s-V_m$ relation (or TFR).

It is important to  note that this scatter is correlated
with some intensive properties. For example, in Figure 5
is seen that, for a given mass, $V_m$ increases
with the $B-V$ color. This is also evident from Table 1 where
we show the correlation coefficients for $A_{\rm TF}\equiv L_i/V_m^{\rm m_i}$
which reflects the deviations from the TFR. This behaviour is easily 
understood from the point of view of the extended collapse 
scenario: galaxies formed through 
gentle MAHs will be less concentrated (smaller $V_m$) and with SF histories more
extended in time (bluer colors) than galaxies formed through early active
MAHs. The observational data confirm this prediction of the models. In
Figure 5, galaxies from a cross of the RC3 (de Vaucouleurs et al. 1991) and
the Tully (Tully 1988) catalogs are also depicted. To estimate the behavior
of $V_m$ with the $B-V$ color for a given mass (luminosity), the data were 
divided into three broad bins according to the $B-$luminosities presented in the 
mentioned catalogs. The dashed lines are lineal regression to each one of these 
bins. Although the estimate is very qualitative, we see that the observed trend 
is in agreement with the models' predictions.


\subsection{The color Tully-Fisher and color-magnitude relations}

The measured slope of the TFR is smaller as the color passband in which the 
luminosity is measured is bluer. This dependence of the TFR slope with the 
color band gives rise to the so called color TFR (Gavazzi 1993; Bothun 
et. al 1985):
\begin{equation}  
\frac{L_i}{L_j}\propto V_m^{(m_i -m_j)},
\end{equation}
\noindent where $i$ and $j$ indicate a given color band, and $m_i$ and $m_j$ are 
the corresponding slopes of the TFR in these bands. For example, for the $H$ and 
$B$ bands, according to the empirical estimates reported in the literature, 
$(m_H-m_B)\approx 0.4-1.2$ (e.g., Gavazzi 1993; 
Strauss \& Willick 1995, and more references therein).
As it was mentioned in the previous subsection, $M_s \propto L_H$. 
Thus, according to eq. (4) we have that 
\begin{equation} 
M_s/L_B \propto V_m^{\beta},
\end{equation}
\noindent where $\beta=m_H-m_B\approx 0.4-1.2$, i.e. the $M_s/L_B$ ratio 
increases with $V_m$. This is an important result which should be seriously
considered in any theory of disk galaxy formation. We find at least three 
possible causes for this dependence of the mass-to-luminosity ratio 
on $V_m$: (i) the SF efficiency could depend on the gravitational 
potential of the system in such a way that in galaxies with larger 
$V_m$ the gas is consumed into stars more efficiently than in galaxies
with low $V_m$; (ii) if disk galaxies build up inside-out by a 
continuous process of gas accretion, then the dependence given by eq. (5)
is expected in the case the rate of gas accretion 
at late epochs decreases more rapidly for larger galaxies than for smaller 
ones\footnote{Such a situation is possible if, for example, in large systems a 
fraction of the accreting gas is trapped and efficiently transformed into stars 
within the potential wells of relative large substructures (satellites) around 
the central disk galaxy.};  (iii) the color TFR or the 
dependence given by eq. (5) may be explained by the effect of internal
dust absorption ---which affects the luminosity in the blue bands but 
not in the infrared ones---, if the absorption increases with 
the galaxy mass (rotation velocity).

The results of our models show that the slopes $m_H$ and $m_B$ 
of the TFRs in the $B-$ and $H-$bands are approximately equal, or, 
what is the same, the $M_s/L_B$ ratio is constant with $V_m$. 
This is (i) because, according to our SF mechanism, the
SF efficiency scarcely depends on mass (or $V_m$), 
and (ii) because we assume that the rate of gas accretion is determined
only by the cosmological rate, avoiding possible intermediate processes
like that mentioned in the footnote. In fact, in our models, at late
epochs the cosmological accretion rates ---given by the MAHs--- for 
larger systems decrease more slowly than for smaller systems (small
halos collapse on average earlier than big halos, Navarro, et al. 1996,
AFH98). However, this dependence on mass (oposite to observations) 
is compensated by the fact that larger galaxies have on average slightly 
higher surface densities (see $\S 3$; Dalcanton et al. 1997; FA), and 
in our models, the SF efficiency is larger for higher surface densities.
It is interesting to note that the same argument was used to explain why
the shape of the rotation curves is nearly independent of mass in spite
of that less massive CDM halos are more concentrated than the
more massive ones ($\S 3$).

We propose that the dependence given by eq. (5) is basically 
produced by dust absorption. It is well known that the metallicity 
abundance of spiral galaxies correlates with their luminosities, 
$Z\propto L_B^{\epsilon}$, where $\epsilon \approx0.3-0.5$ (see references in 
Roberts \& Haynes 1994). Since dust particles form from heavy elements,
this relation suggests that more luminous galaxies have higher dust 
abundance than the less luminous galaxies. Direct galaxy images also
show that the dust abundances increase with the galaxy
luminosity (van den Bergh \& Pierce 1990). Indeed, WH96 based on 
studies of the UV and FIR fluxes of a sample of normal late-type 
galaxies, have found a clear correlation between the dust opacity 
and the $B-$band luminosity of galaxies.

In the Appendix, from the results of WH96 and using the 
uniform slab model, we calculate the extinction in the $B-$band,
$A_B$, as a function of $L_B$ (eq. A2). Applying
this luminosity-dependent extinction to our models we obtain that the 
slope of the TFR in the $B-$band becomes flatter than in the original models.
Figure 6 shows that this relation in the range 
log$L_B/L_{B_{\odot }}\approx 9.5-11.5$ is approximated 
by a line with slope $\sim $2.7, as many observations 
suggest. Note that the dependence of extinction on luminosity 
does not only produce a change of slope, but
also some nonlinearity, particularly at the bright end of Figure 6. A
similar result was previously reported by Giovanelli et al. (1995). If
the samples used by different observers cover different luminosity ranges,
this could explain why the reported slopes of the $B$-band TFRs are so 
disparated. Recently, Kudrya et al. (1997) have presented the $B-$band 
TFR for a large sample of galaxies in a large range of
luminosities; from their Figure 6 it is clearly seen how the slope of this 
relation tends to be steeper for the less luminous galaxies 
(see also e.g., Pierce \& Tully 1992). In 
Figure 6 are also depicted the predicted TFRs after
applying extinctions calculated with the maximal and
minimal optical depths given in WH96 (see Appendix). 
The more realistic ``sandwich'' model (Disney, Davies, \& Philipps 1989)
was also considered (dotted line, see Appendix).

Since the early eighties a correlation between the galaxy total 
color and its magnitude was reported for spiral galaxies 
(Vishnavatan 1981; Wyse 1982; Tully, Mould, \& Aaranson 1982). 
The more luminous is the galaxy the redder tends to be its color.
As a matter of fact, this relation is intimately related to the color
TFR mentioned above. For instance, by dividing the $B-$band 
$L_B=C_BV_{m }^{m_B}$ and $H-$band $L_H=C_HV_{m }^{m_H}$ TFR 
one to another, one obtains:
\begin{equation}  
(B-H)=2.5(\frac{m_H}{m_B}-1)\log L_B+2.5\log \left( \frac{C_H}{C_B^{m_H/m_B}}%
\frac{L_{B_{\odot }}}{L_{H_{\odot }}}\right) 
\end{equation}
For the same range of values for $m_B$ and $m_H$ used in eq. (5), 
we have that $(B-H) \propto \alpha \log L_B$ with 
$\alpha \approx 0.4-1.3$. This is in rough agreement with the 
slopes reported for the color-magnitude relation. Thus, according
to the above discussion, dust absorption may be responsible 
for this relation. 
Assuming that in the $H-$band dust absoption is negligible, 
while in the $B-$band absorption depends upon luminosity
as WH96 found (eq. A1), the galaxy $(B-H)$ color will redden 
proportionally to the extinction $A_B$ estimated in the Appendix. As a first 
approximation, using the linear fitting to $A_B$ (eq. A3), one obtains that 
\begin{equation} 
(B-H)\propto 0.42\log L_B.
\end{equation}


WH96 suggested that the dependence 
of internal absorption upon luminosity (eq. A1)  may be explained 
by the observed increase of metallicity and/or surface density with
the luminosity. Less luminous (massive) galaxies can have lower 
metallicity abundances than more luminous galaxies because the 
dust expelling from the galaxy is more effective in the former than
in the latter. Shustov, Wiebe, \& Tutukov (1997), using models 
of disk galaxy evolution that take into account the process of dust
expelling by radiation pressure, show that dust loss is much
more effective in low-mass galaxies than in massive ones.
They obtained a dependence of iron abundance 
on galaxy luminosity near to the observed, and found that 
the other properties of the galaxy models do not significantly
change with dust loss. Hydrodynamical
simulations of starburst-driven mass ejection indeed have shown
that the efficiency of metal ejection is much larger than that of the
gas and this efficiency of metal ejection decreases with mass 
(Mac Low \& Ferrara 1999). Regarding the disk surface density, our
models indeed show that more massive galaxies have typically 
slightly higher surface density disks. Although this dependence is not 
too pronounced as the one Dalcanton et al. (1997) obtained, it could also 
contributes to give rise to the absorption-luminosity relation (eq. A1).

Finally, we note that metallicity, in addition to the absorption effect, 
also influences the spectrophotometric evolution. For example the 
spectrophotometric  models of 
Bressan, Chiosi, \& Fagotto (1994) applied to single stellar populations with
different initial metallicities show that at $\sim 12$ Gyr the differences
in the $V-K$ and $B-V$ colors are $\sim $0.88 mag and $\sim $0.23 mag,
respectively for a factor of 20 of variance in metallicity. These models
roughly agree with the observed color-magnitude relation of elliptical
galaxies, and in accordance with Kodama \& Arimoto (1997), this relation is
consequence of a metallicity effect, instead of an age effect. 

\subsection{ The luminosity-radius relation}

The mass-radius relation predicted by the models (the average values in
the MAH and $\lambda $ were used) is:
\begin{equation} 
M_s\propto R_H^{2.3},
\end{equation}
where $M_s$ is the disk stellar mass, and $R_H$ is the Holmberg radius. The
radius scales with $L_B$ as $R_H^{2.4}.$ The scatter
in these relations is large and it correlates with the SB
(AF99). The Holmberg and scale radii do not correlate
with any intensive galaxy property (see Table 1), suggesting that the
evolution of disk galaxies and the Hubble sequence are size independent 
(de Jong 1996a; McGaugh \& de Block 1997).

\section{The fundamental physical factors of disk galaxies and the Hubble
sequence}

One of the aims of this paper is to investigate which physical
factors determine the properties of present-day disk
galaxies. According to the extended collapse scenario, 
these fundamental factors are: {\it  the virial present-day mass 
$M_v$ of the system, the MAH, and the angular momentum} expressed 
through the spin parameter $\lambda $. These factors and their 
statistical distributions are related to the initial cosmological 
conditions and they are able to produce most of the disk galaxy
properties and their correlations. As a first approximation, here 
we assumeed independence between $\lambda $ and the MAH. Analytical 
studies (c.f., Hoffman 1986, 1988; Heavens \& Peacock 1988) 
suggest that $\lambda $ is almost independent from the 
fluctuation peak height (related to the MAH) for CDM
power spectra. 
We also assumed independence between $\lambda $ and $M_v$ (see \S 2.3).
Concerning the dependence of MAH upon $M_v$, the less
massive galaxies undergo a faster early collapse than the 
more massive galaxies for the CDM power spectra (AFH98). 

The correlation coefficients from a principal analysis of these three
factors with the model galaxy properties are given in Table 2.
The MAH was quantified through the $\gamma $ parameter, where 
\begin{equation}
\gamma =\frac{log(M_{\rm nom}/(M_{\rm nom}/2))}
{log(t(M_{\rm nom})/t(M_{\rm nom}/2))}.  
\end{equation}
The MAH, which partially drives the SF
history, strongly influences the $B-V$ color index, and moderately influences 
the gas fraction $f_g$. The range of $B-V$ colors that the models span is mainly 
associated with the statistical dispersion in the MAHs. This dispersion is also
reflected in the TFR scatter (the coefficient $A_{\rm TF}$ correlates with
$\gamma$), and as it was shown above (see Figure 5),
observations confirm a correlation between $B-V$ and the maximum circular
velocity for a given luminosity (mass). 

The $\lambda $ parameter strongly influences the SB, the b/d ratio, and $f_g$. 
For a given $M_v$ and MAH, $\lambda $ determines the degree of 
concentration (SB) of the disks. According to the SF 
mechanism used in
our models, the surface density influences the SF efficiency and 
therefore, the $f_g$. Furthermore, as it was mentioned in $\S$ 4, 
the stellar Toomre parameter $Q_s$ is small for small stellar surface
densities. Therefore, the b/d ratio is larger for systems with 
small $\lambda^{\prime }s$.

Mass strongly correlates with luminosity and the Holmberg radius and
slightly influences the b/d ratio and the SB (both in the
same direction as suggested by observations). As it was pointed out in \S 3.2,
$B-V$ and $\mu _{B_o}$ are the two parameters on which the other intensive
properties depend, i.e. the intensive properties of disk galaxies may be
described in a biparametrical sequence, whose origin deals with two of the
fundamental physical factors of galaxies, the MAH and $\lambda ,$
respectively. Moreover, we have found that these properties correlate between 
them following the same trends that observed galaxies present across
the Hubble sequence. Most of the intensive properties are almost independent from 
the third fundamental factor, the mass (luminosity). We have found that 
the mass-dependent dust absorption may be responsible for the 
color TF and color-magnitude relations ($\S 5.2$).


The morphological Hubble classification has been a useful guide to study the
observational properties and correlations of galaxies. Nevertheless, most of
the classification criteria deal with morphological characteristics
that probably are transient phenomena related to more fundamental
galaxy properties. The main classification criteria for spirals
are the pitch angle and the strength of the spiral arms, and the b/d ratio. 
Despite the big effort done, the problem of the origin and maintenance 
of arms in disk galaxies is still not well understood.
Bertin \& Romeo (1987), Bertin et al. (1989a,b) and other authors (see for
references Combes 1993) coincide in pointing out that gas, dynamically
speaking, is crucial for the excitation and maintenance of spiral
structures.
According to Bertin \& Lin (1996), the gas fraction mainly determines the
sequence of types a, b, and c, i.e. the pitch angle of arms. As is seen in
Table 1, the gas fractions in our models correlate with the other secondary
indicators of the Hubble sequence which go across the a, b, and c sequence, 
such as b/d, $\mu_{B_0},$ and $B-V$, suggesting that the origin of 
the Hubble morphological types is closely related to the galaxy 
formation and evolution processes of the extended collapse scenario.

Several observers have pointed out that, for a given luminosity, galaxies 
with larger maximum rotation velocities (the TFR scatter) are of earlier 
types than those with smaller velocities (Roberts 1978; Rubin, 
Thonnard, \& Ford 1980, Rubin et al. 1985; Giraud 1986, 1987; 
Krann-Korteweg, Cameron, \& Tamman 1988; Giovanelli et al. 1997). The TFR, 
$L_{i}=A_{\rm TF}V_{m}^{m_i},$ where $i$ is a given band,
predicted by the models has a dispersion that, at a first approximation, we
express only through variations in the coefficient $A_{\rm TF}$. As has been
commented in \S\ 3.2 (see also Fig. 5),  the models
present correlations between $A_{\rm TF}$ and some secondary Hubble type
indicators as $B-V$, $f_g,$ and b/d (Table 1). Thus, within the framework
of the extended collapse scenario, the empirical correlation found between
galaxy type and rotation velocity for a given luminosity finds a natural
explanation.

\section{Discussion}

\subsection{Problems of the hierarchical formation scenario}

We have studied disk galaxy formation and evolution in the cosmological 
context taking into account several key factors and processes of this
phenomenon. Our models represent an extreme case of galaxy formation 
where we have assumed: (i) gentle and spherical symmetric gas infall 
at the center of hierarchically growing CDM halos (no mergers),
(ii) no gas reheated and expelled back into the halo or 
lost from the system, and (iii) stationary self-regulated SF 
{\it within} the disk. We have seen
that at the level of this simple picture many properties and correlations of the
observed disk galaxies were predicted. However, the models are
in potential conflict with some observational pieces of evidence.
As we shall see, these conflicts may be partially solved invoking some
extra ingredients in our scenario, but it is also possible that they are
pointing out to serious inconsistences of the hierarchical clustering
scenario in general.

In an accompanying paper (Firmani \& Avila-Reese 2000b) we study the evolution
of the models presented here. We find that the size and SB evolution of the 
disks is more dramatic than that one the deep field observations seem to suggest.
This problem, toghether with the too steep negative radial color gradients
our models predict ($\S 3$), constitute serious difficulties for the extended
collapse scenario. The color gradients we obtain might be even steeper if the 
galaxy's internal extinction significantly depends on surface density. 
The model disks also tend to be somewhat less dense and more 
extended than real galaxy disks, and the gas fractions of the 
models are slightly larger than those measured in galaxies (see Figs. 3 and 4). 
On the other hand, the integral $B-V$ colors of observed red galaxies
seem to be redder than the red models (Figs. 3 and 4). In $\S 4$ we have
extensively discussed the solutions to the last conflict.

If some transfer of angular momentum from baryon matter to DM is 
allowed, then the disks result smaller and more
concentrated. If this transfer of angular momentum tends to be
more important at latter epochs and/or if a fraction of the late, 
high angular momentum
accreting gas is trapped within substructures and does not fall 
to the disk, then the color gradient will be smaller, the total color 
index will be redder, and the size evolution will be less pronounced 
than we have obtained. It is also possible that during  the gas 
collapse, the original 
gas angular momentum is redistributed due to several dynamical and hydrodynamical
processes related to the deviations from the homogeneity and sphericity. 
Experiments where a moderate ``flattening'' of the radial angular momentum 
distribution of the infalling gas was introduced,  show us that the disk radial 
color index gradients significantly decrease and the disk 
sizes change more slowly in time with respect to the original models.

It is interesting to note that the above mentioned conflicts of 
the extended collapse
scenario tend to be opossed to those of the merger scenario. The 
merging seem to cause serious difficulties for conserving the gas initial angular 
momentum, with the consequence that disks result too small. This
problem, called the ``angular momentum catastrophe'',  has been 
observed in the N-body+hydrodynamical numerical simulations without SF (e.g., 
Navarro \& Steinmetz 1997). When SF is included, the merging of the large
stellar subunits as the system evolves, also produces loss of angular momentum
with the consequent formation of a elliptical-like object instead of a disk
(c.f., Katz 1992). It is probable that reality lies 
in between the extended collapse and merger scenarios. More  
observational and theoretical studies on the formation and evolution 
of disks, their SF histories, and the radial distribution of color 
indexes and extinction are needed in order to arrive 
at definitive conclusions.

While some physical ingredients not considered in our scenario 
actually might work in the direction of improving the predictions, 
we do not discard the possibility that
in general the hierarchical formation picture is in conflict with reality. 
On the other hand, as already has been pointed out, if the merging activity 
in the hierarchical formation picture based on the Gaussian inflationary CDM 
models is actually too high, then it is difficult to figure out the existence 
of a large number of galaxy disks in the framework of this picture 
(e.g., T\'oth \& Ostriker 1992). 
To the problems related with the formation of disks in the standard hierarchical
formation picture, we add other potential conflict: the inner 
structure of the CDM halos. This is the subject of the next subsection.

\subsection{Effects of a shallow core in the DM halo}

The rotation curves of dwarf and low SB galaxies suggest that the
inner density profiles of DM halos are shallower than $r^{-1}$ 
(Moore 1994; Flores \& Primack 1994; Burkert 1995; de Blok \& McGaugh 1997). 
Our results and the results of high-resolution cosmological N-body 
simulations (e.g., Navarro et al. 1996,1997; Fukushige \& Makino 1997; 
Moore et al. 1998, 1999; Jing 1999; Jing \& Suto 1999) show that for 
the CDM models the halo inner density profiles are not shallower 
than $r^{-1}$. The rotation curve decomposition of the galaxies
formed within the CDM halos also seem to be in conflict with the 
decompositions inferred from observations for high SB galaxies (AFH98; FA;
but see Navarro 1998). 

 In order to explore the effects of a shallow core on the
properties of the model disk galaxies, we introduce an artificial core in our
evolving CDM halos. Our approach is purely phenomenological: the inner density
profile of the evolving halos is flattened in such a way that the final rotation 
curves of low
SB model galaxies (those with large $\lambda$) resemble the observed rotation
curves of low SB (see for details FA). The inner 
halo density profile obtained for the SCDM $\sigma _8=0.6$ model treated here is 
very similar to that obtained for the $\Lambda$CDM model in 
FA. 

As in FA, we find that the main effect of the shallow core is to raise the 
contribution of the disk component with respect to the halo component
in the rotation curve decomposition. This makes the rotation curve
decompositions of the model galaxies more realistic.
The intensive properties are almost the same for 
the models with and without a shallow core. The $V_{m}$ of the 
models with a shallow core are slightly smaller than for the original
ones. This shifts our predicted $I-$ and $H-$band TFRs to lower velocities
thus giving a better agreement with the observational data (see also FA). 
For a fixed rotation curve shape, the central surface densitites (SB) 
are higher for the models with core. This is because in order to 
have the same rotation curve shape, a model with a shallow core 
should have a more concentrated disk (smaller $\lambda$) that 
would compensate the smaller halo contribution to the total rotation
curve with respect to the model with a more dense halo core. 

\section{Summary and conclusions}

We studied the formation and evolution of 
individual disk galaxies within the framework of an inflationary 
Gaussian SCDM model normalized to $\sigma _8=0.6$. We built-up 
disks in centrifugal equilibrium within hierarchically growing 
DM halos; the disks form inside-out with gas (no mergers) accreted
with a rate proportional to the cosmological infall rate given by 
the mass aggregation history (MAH), and under the assumption 
$\lambda =const$ in time and 
detailed angular momentum conservation during the gas collapse. The 
SF in the disk is produced by global gravitational
instabilities and it self-regulates by an energy balance in the turbulent
ISM. We calculated the secular formation of a bulge using a gravitational 
instability criterion for the stellar disk. Our main conclusions are:

1). The main properties of present-day high and low SB disk galaxies and 
their correlations may principally be explained by the combination of three 
fundamental physical factors and their statistical distributions. These factors
are the halo virial mass $M_v$, the MAH, and the spin parameter $\lambda$,
and they are related to the cosmological initial conditions.

2). The SF efficiency of the models does not sustantially depend 
upon the mass or $V_m$
of the galaxy. Therefore, the integral color index $B-V$ or  
the mass-to-luminosity ratio 
$M/L_B$ do not depend upon mass. We showed that the empirical correlation 
of dust extinction upon luminosity reported by WH96 explains the 
observed color-magnitud  and  color TFRs without the necessity
of introducing a mass-dependent SF efficiency. 

3). Several global and local properties of present-day disk galaxies and their 
correlations are a natural consequence of the extended collapse
scenario. We remark the following results:

- The redder and more concentrated is the model disk, the smaller is its gas 
fraction and the larger is its b/d ratio. The values of these 
intensive properties are in rough agreement with the observational 
data. The intensive properties of the models
are described by a biparametrical sequence, where the parameters are
the color index $B-V$ and the central SB, $\mu _{B_o}$. 
These parameters are
mainly determined by the MAH and by $\lambda $, respectively. Thus, according
to our models, the Hubble sequence is biparametrical as some observations suggest.  

-The model results agree with the near-infrared TFRs, suggesting that the origin 
of the TFR is cosmological. This relation is almost independent
from the SB and its scatter is mainly dominated by the dispersion in the MAHs 
(or the dispersion in the DM halo structures). For a given luminosity, the
larger the $V_m$ is (TFR deviation), the redder is the galaxy and the smaller 
is its gas fraction $f_g$. This is because, on one hand the MAH influences on 
the galaxy color index and $f_g$, and on the other hand, it is 
the dispersion in the MAHs that mainly produces the deviations from the TFR. 

-The stellar surface density and $B-$brightness profiles are nearly exponential.
This is a result of the inside-out disk formation assuming detailed angular 
momentum conservation and $\lambda$ constant in time. The SB of the
models mainly depends on $\lambda$.

-Also as a result of the inside-out disk formation, the radial $B-V$ 
gradients  along the disks are negative: the outer disk 
regions are younger than the inner ones. 

-The shape of the total rotation curves depends on the the disk mass 
fraction $f_d$, $\lambda$ and the MAH. For reasonable values of $f_d$, 
the approximate flatness of the rotation curves presented by most of the 
models is due to a ``conspiration'' among the CDM halo and the baryon disk. 
The rotation curve decompositions, however, show excesive dominance of the halo
component over the disk component with respect to observational estimates.

 The main shortcomings of the models presented here are (i) the probably 
excesive radial color 
index gradients and disk size evolution,  and (ii) the DM dominion in the 
rotation curve decompositon. The inclusion in the models of other,
less important ingredients and processes
as merging, angular momentum transfer, and non-stationary SF work in the 
direction of overcoming the former shortcomings. The latter problem 
is solved if the inner density profile of the CDM is shallower than 
predicted. It is probable that
these problems are in general pointing to a more serious difficulty
of the hierarchical formation scenario based on the Gaussian CDM models. 

\acknowledgments
{\bf Acknowledgments}

We are grateful to the referee for helpful comments.

\appendix

\section{Calculation of the luminosity-dependent $B-$band exctinction}

Wang \& Heckman (1996), based on studies of the far UV and
FIR fluxes of a sample of normal late type galaxies, have concluded that the
dust opacity increases with the luminosity of the young stellar population.
They find that a power law relation between the optical depth and the UV 
luminosity explains
this observational dependence. This relation, referred to the B band, is:
\begin{equation}
\tau _B=\tau _{B,*}\left( \frac{L_B}{L_{B,*}}\right) ^\beta 
\end{equation}
where the best fits to observations are for $L_{B,*}=1.3\times
10^{10}L_{B_{\odot }},$ $\tau _{B,*}=0.8\pm 0.3,$ and $\beta =0.5\pm 0.2$.
According to the uniform slab model WH96 
have used, the extinction in magnitudes may be expressed as 
$A_B=-2.5\log \left( \frac{1-\exp (-\tau _B)}{\tau _B}\right) $, and 
in the range $10^8-10^{11}L_{B_{\odot }},$ using (A1)
with the central values, is well approximated by 
\begin{equation}
A_B\approx 0.38+0.42\log 
\left( \frac{L_B}{10^{10}L_{B_{\odot }}}\right) +0.14\left( \log \left( 
\frac{L_B}{10^{10}L_{B_{\odot }}}\right) \right) ^2
\end{equation}
In the range of $10^9-10^{11}L_{B_{\odot }}$ the linear fitting
\begin{equation}
A_B\approx 0.43+0.42\log \left( \frac{L_B}{10^{10}L_{B_{\odot }}}\right) 
\end{equation}
is also a good approximation.

To calculate the extinction for the two extreme cases of maximal and
minimal optical depths, we use $\tau _{B,*}=1.1$ and $\beta =0.7,$
for the former, and $\tau _{B,*}=0.5$ and $\beta =0.3$ for the latter. 
To calculate the extinction with the more realistic ``sandwich'' 
model (Disney, Davies, \& Philipps 1989), we use the fiducial value
of $\beta $ given by WH96, and $\tau _{B,*}$ and the 
ratio of the height scale of dust to young stars, $\zeta ,$ are fixed to 
the values suggested by Bosselli \& Gavazzi (1994) for the optically 
thin case in the $H$ band ($\tau _{B,*}=$1.33 and $\zeta =$0.74).

\clearpage

\begin{figure}
\vspace{17cm}
\includegraphics{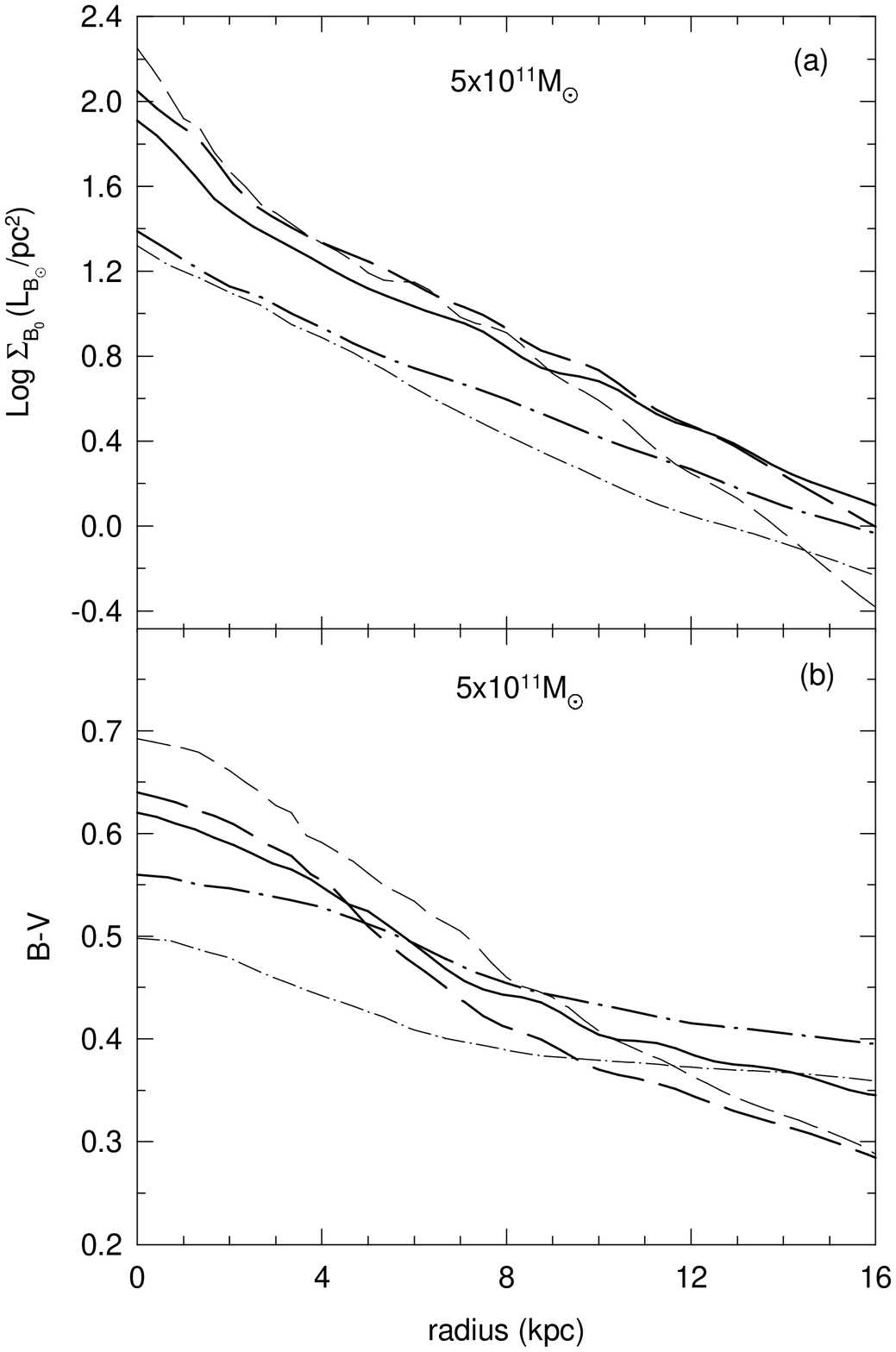}
\caption[fig01.eps]{The $B$-surface brightness (a) and the $B-V$ color index
  (b) profiles of a 5$\times 10^{11}M_{\odot }$ galaxy. The average
  MAH cases for the spin parameters $\lambda =0.035$ (dashed line),
  $\lambda =0.050$ (solid line), and $\lambda =0.100$ (point-dashed
  line) are represented with the thick lines. For the early active MAH
  only the model with $\lambda =0.035 $ (thin dashed line) is plotted,
  while for the extended MAH, only the $\lambda =0.100$ case (thin
  point-dashed line) is plotted. All the models were calculated for a
  $\sigma _8=0.6$ SCDM\ model. \label{fig01}}
\end{figure}

\begin{figure}
\vspace{8cm}
\includegraphics{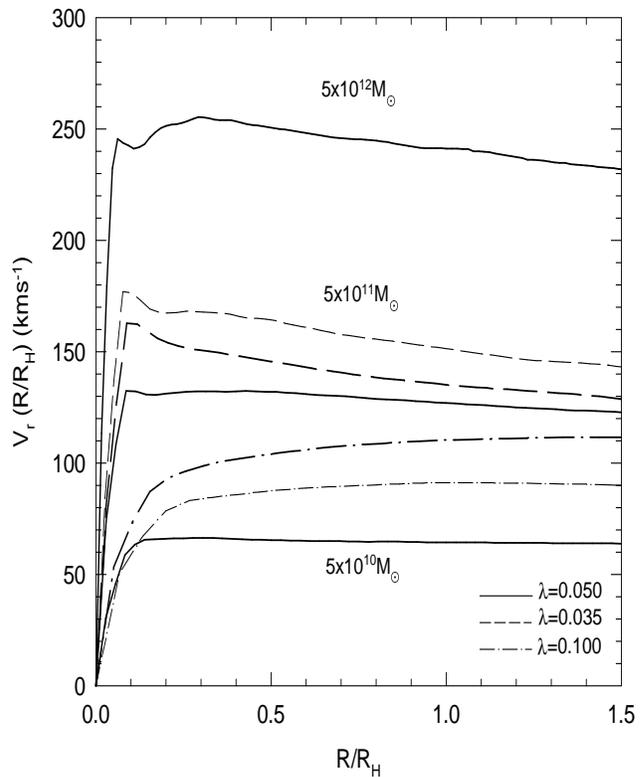}
\caption[fig02.eps]{Rotation curves for the same models of Figure 1 
(5$\times 10^{11}M_{\odot })$, and for a 5$\times 10^{10}M_{\odot }$ (bottom
  curve) and 5$\times 10^{12}M_{\odot }$ (top curve) galaxy
  corresponding only to the average MAH, $\lambda =0.050.$ Radii were
  scaled to the optical (Holmberg) radii of each model. \label{fig02}}
\end{figure}


\begin{figure*}
\vspace{11.5cm}
\includegraphics{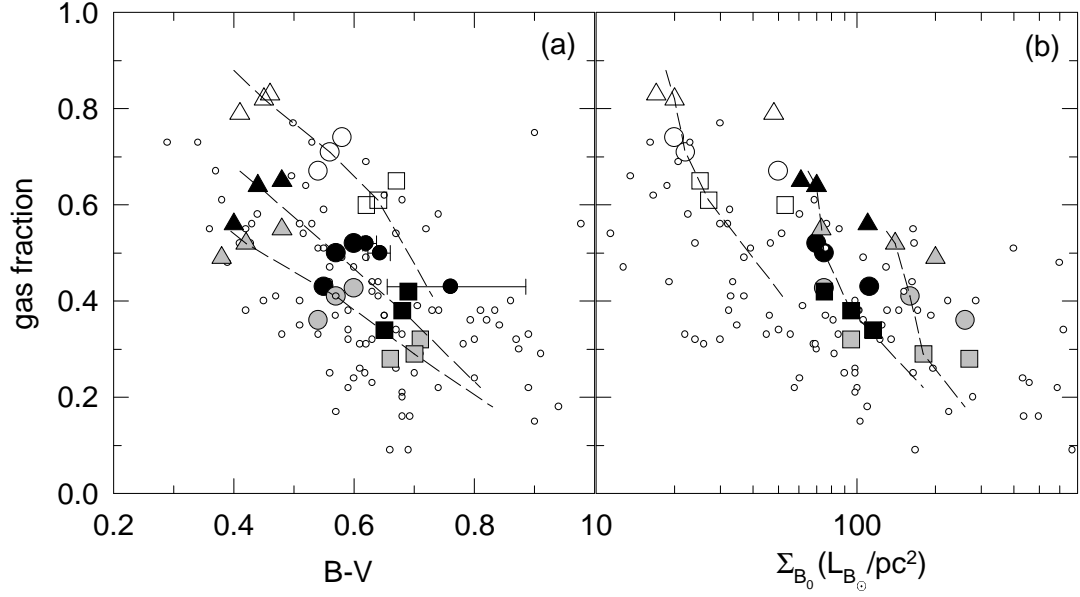}
\caption[fig04.eps]{The gas fraction $f_g$ vs. the integral $B-V$ color 
index (a), and vs. the central $B$-SB $\mu _{B_0}$ (b) for
  models and observations. The gray, black, and white filled symbols,
  correspond to models with $\lambda =0.035,$ $\lambda =0.050,$ and
  $\lambda =0.100$ respectively. Squares are for the early active MAH,
  circles for the average
MAH, and triangles for the extended MAH. Three masses (dark+baryon), 5$%
\times 10^{10}M_{\odot },$ 5$\times 10^{11}M_{\odot },$ and 5$\times
10^{12}M_{\odot }$ are considered (the larger the mass, the smaller is
the
gas fraction). The dashed lines connect the models of constant mass for 5$%
\times 10^{11}M_{\odot },$ and extend the statistical range of MAHs to
94\% (symbols consider only 80\% of the MAHs). The three small black
filled circles are the same models corresponding to the big black
filled circles but reddened according to the dust
absorption-luminosity dependence given in WH96 (see
text). The error bars correspond to the range of values given 
in WH96. Small empty circles are the observational data
collected by McGaugh \& de Blok (1997) and corrected for inclination
extinction. LSB\ galaxies are included. \label{fig03}}
\end{figure*}

\begin{figure*}
\vspace{14cm}
\includegraphics{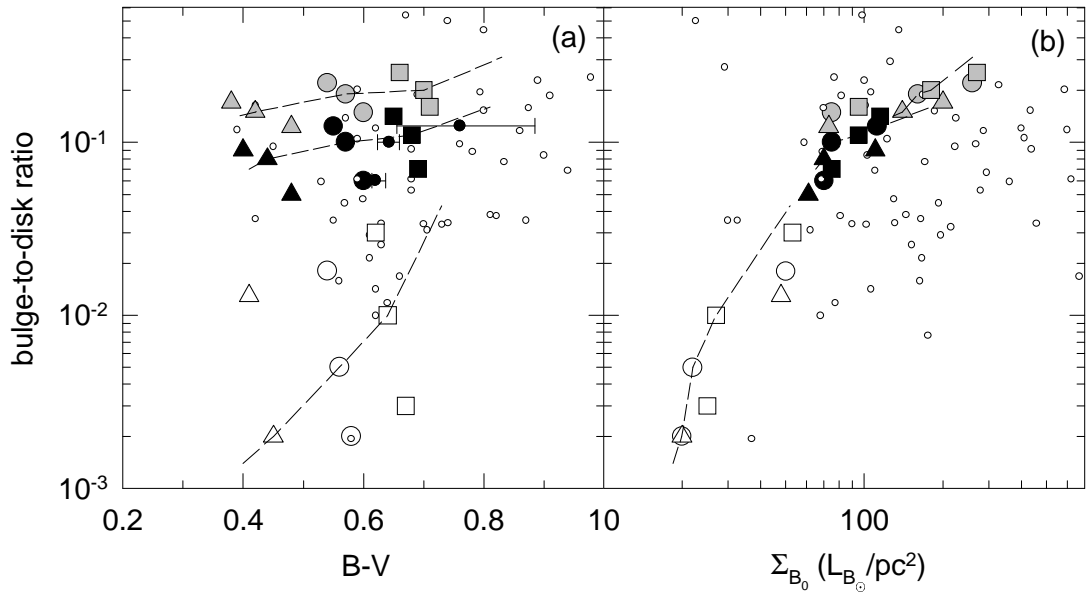}
\caption[fig05.eps]{The b/d ratio vs the integral $B-V$ color index
  (a), and the central $B$-SB $\mu _{B_0}$ (b) for
  models and observations. The same symbol and line codes of Figure 3
  are used. The bulge-to-disk ratios were taken from the $K-$band
  two-dimensional decompositions carried out by de Jong (1996b). LSB
  galaxies and a few normal galaxies shown in Figure 4 are absent in
  this Figure. \label{fig04}}
\end{figure*}

\begin{figure*}
\vspace{14cm}
\includegraphics{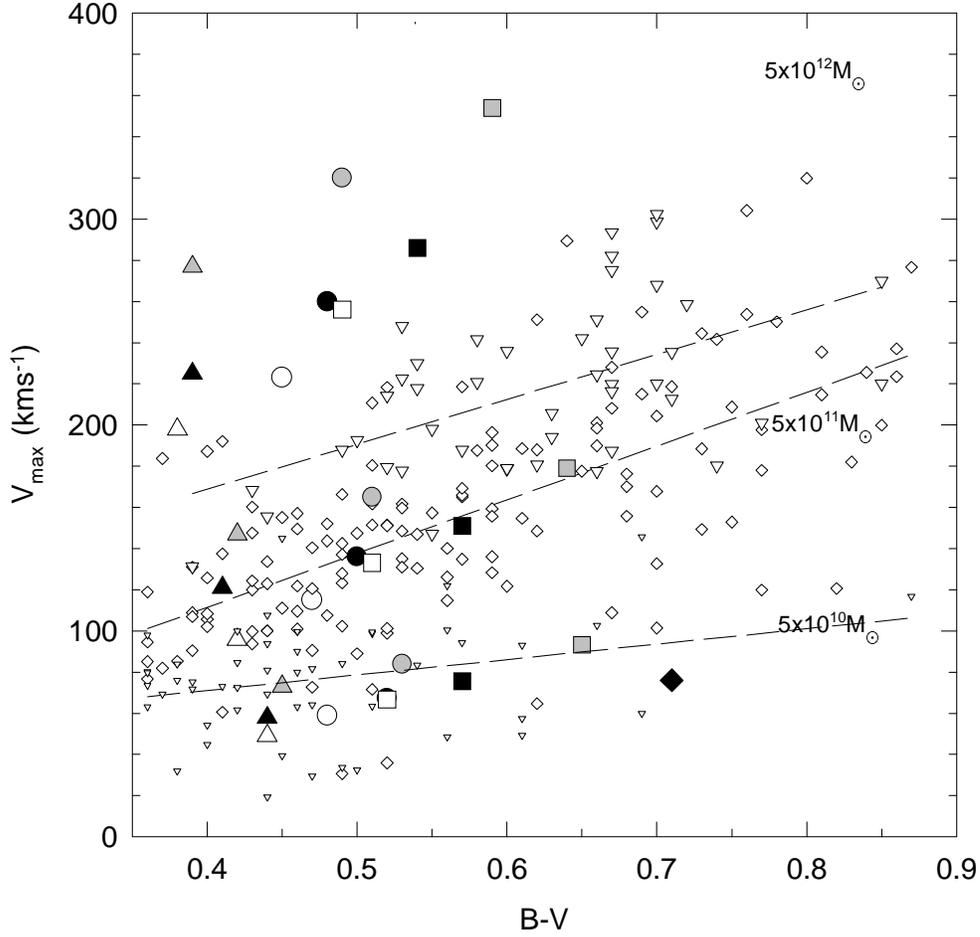}
\caption[fig06.eps]{The maximum rotation velocity, $V_m$ vs. $B-V$ 
for models and
  observations. The same symbol codes of Figure 3 are used. The
  observational data (small symbols) were taken from a cross of the
  RC3 and the Tully (1988) catalogs (see text). The small triangles,
  diamonds, and inverse triangles
correspond to galaxies with luminosities in $B$ band within the 10$^8-3%
\times ${10}$^9L_{B_{\odot }},$ 3$\times ${10}$^9-3\times
10^{10}L_{B_{\odot }}$, and 3$\times 10^{10}-2\times
10^{11}L_{B_{\odot }}$ ranges, respectively. The dashed lines are
linear regressions to the observational data corresponding to these
ranges. Note how the maximum velocity of models and observations for a
given mass (or range of luminosities) correlates 
with the $B-V$ color. \label{fig05}}
\end{figure*}

\begin{figure*}
\vspace{14cm}
\includegraphics{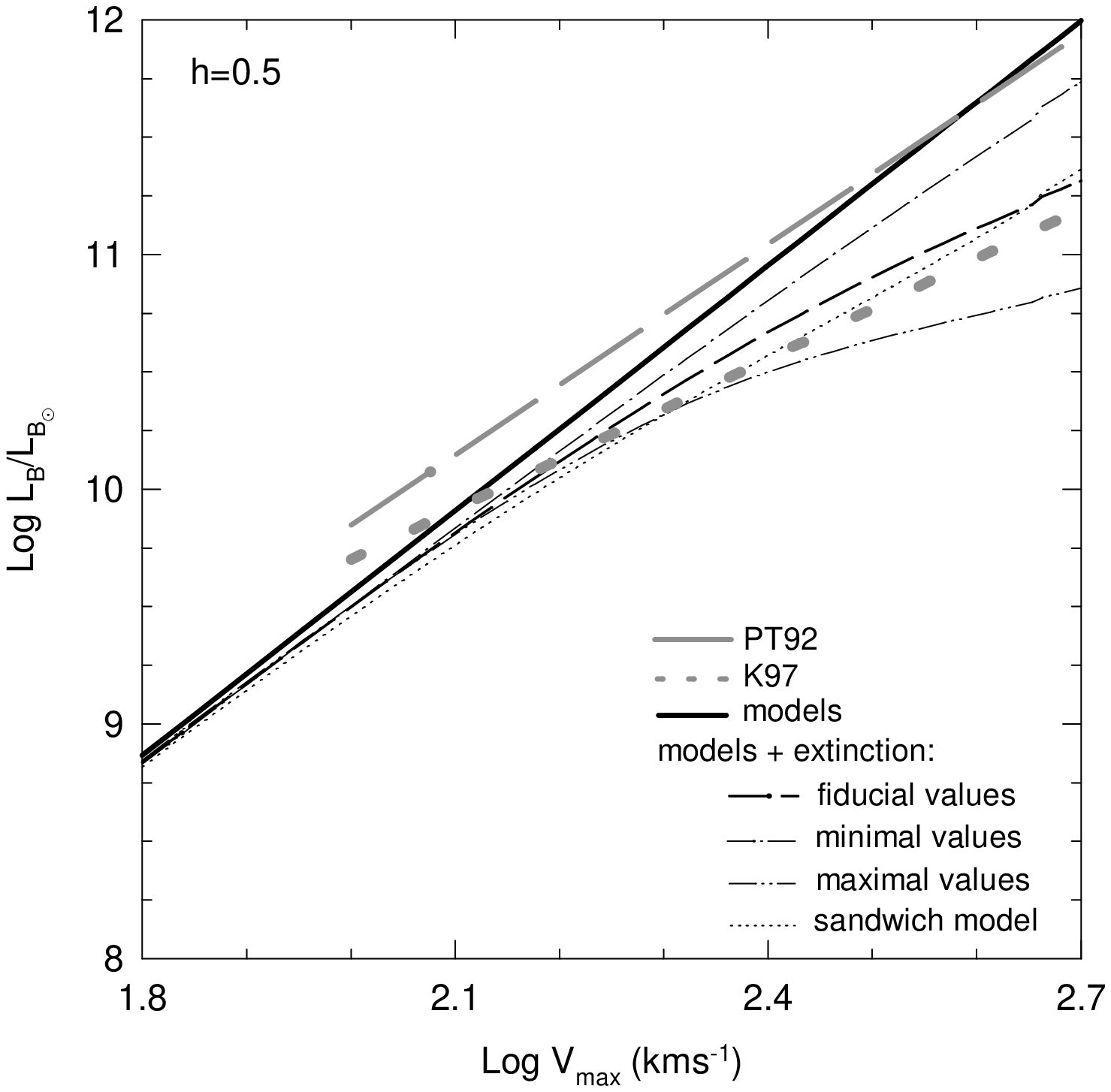}
\caption[fig07.eps]{The predicted $B$-band TFR for the 
$\sigma _8=0.6$
  SCDM model (thick solid line). The slope of this relation is 3.5.
  The other black lines show how the intrinsical TFR is transformed if
  the $B$-luminosities are dimished by dust absorption according to the
  observational dependence of optical depth of dust on luminosity
  given in WH96. While the dashed line corresponds
  to the fiducial optical depth, the point-dashed and two point-dashed
  lines are for two extreme cases of maximal and minimal optical
  depths (see text). The point line was obtained using a sandwich
  model with a ratio of height scale of dust to young stars of 0.74
  (see appendix for references). In the range
  log$L_B/L_{B_{\odot }}\approx 9.5-11.5$
  the dashed line is well approximated by a line with slope $\sim
  2.7$. The dotted and dashed gray curves are the empirical TFRs
  given in Kudrya et al. (1997) and Pierce \& Tully (1992), respectively. 
  We have assumed $W_{50}^{corr}=2\times V_{\max }$. These TFRs 
  have a steeper slope for velocities lower 
  than $V_{\max }\approx 100$ km/s (see Figure 6 of Kudrya et al.
  and Fig. 3 of Pierce \& Tully).\label{fig06}}
\end{figure*}

\clearpage

\begin{table}
\caption{Correlation matrix of the global properties \label{tbl-1}}
\begin{center}
\begin{tabular}{lrrrrrrr}\hline\hline
\nodata & B$$-$$V & $f_g$ & b/d & $M_B$ & $V_m$ & log
    $h_d$ & $A_{TF}^{(a)}$ \\ \hline
$\mu_{B_0}$ & $-$0.07& 0.84& $-$0.93& 0.55& $-$0.65& 0.29& 0.33 \\ B$$-$$V & \nodata &
$-$0.56& 0.18& 0.22& $-$0.05& $-$0.29& $-$0.82 \\ $f_g$ & \nodata &
\nodata & $-$0.87& 0.27& $-$0.45& $-$0.01& 0.68 \\ b/d & \nodata &
\nodata & \nodata & $-$0.38& 0.54& 0.10& $-$0.51 \\ $M_B$ & \nodata &
\nodata & \nodata & \nodata & $-$0.94& $-$0.95& $-$0.20 \\ $V_m$ &
\nodata & \nodata & \nodata & \nodata & \nodata & 0.84& $-$0.04 \\
log $h_d$ & \nodata & \nodata & \nodata & \nodata & \nodata & \nodata
& 0.34 \\ \hline\hline
\end{tabular}
\end{center}
{(a) }{$A_{\rm TF}=L_i/V_m^{\rm m_i}$ reflects the deviations
of the models from the TFR}
\end{table}


\begin{table}
\caption{Correlation matrix of the global properties 
and the fundamental parameters \label{tbl-2}}
\begin{center}
\begin{tabular}{lrrrrrrrr}\hline\hline
\nodata          & $\mu_{B_0}$ &
B$-$V            & $f_g$       &
b/d              & $M_B$       &
$V_m$            & log $h_d$   &
$A_{TF}$\\ \hline

$\log M_0^{(a)}$ & $-$0.44 & $-$0.24 & $-$0.17 &    0.27 & 
$-$0.9 & 0.92 & 0.98 & 0.25 \\
$\gamma^{(b)}$  &    0.01 & $-$0.95 &    0.48 & $-$0.12 & 
$-$0.24& 0.05 & 0.29 & 0.82 \\
$\lambda ^{(c)}$ &    0.86 & $-$0.07 &    0.81 & $-$0.88 &   
0.13& $-$0.23 & 0.51 & 0.34 \\ \hline\hline
\end{tabular}
\end{center}
{(a) }{nominal present-day mass} \\
{(b) }{MAH parameter (see eq. 9)} \\
{(c) }{spin parameter}
\end{table}


\begin{thebibliography}


\bibitem{} Andredakis, Y.C., Peletier, R.F., \& Balcells, M. 1995, \mnras,  275, 874

\bibitem{} Andredakis, Y.C., \& Sanders, R.H. 1994, \mnras, 267, 283

\bibitem{}  Avila-Reese, V. 1998, PhD. Thesis, U.N.A.M. (in Spanish).

\bibitem{} Avila-Reese, V., \& V\'azquez-Semadeni 2000, in "Astrophysical Plasmas: 
Codes, Models and Observations", Eds. J.Franco, J.Arthur, N.Brickhouse, 
Rev.Mex.AA Conf. Series, in press

\bibitem{}  Avila-Reese, V., Firmani, C., \& Hern\'{a}ndez, X. 1997, \apj, 505,
37 (AFH98)

\bibitem{}  Avila-Reese, V., Firmani, C., Klypin, A., \& Kravtsov, A. 1999, 
\mnras, 310, 527 

\bibitem{}  Baugh, C.M., Cole, S., \& Frenk, C.S. 1996, \mnras, 283, 1361

\bibitem{}  Baugh, C.M., Cole, S., \& Frenk, C.S., \& Lacey, C. 1997, \apj, 498, 504

\bibitem{}  Begeman, K. 1987, Ph.D. Thesis, University of Groningen

\bibitem{}  Bertin, G., \& Lin, C.C. 1996, ``Spiral Structure in Galaxies. A
Density Wave Theory'' (The MIT Press)

\bibitem{}  Bertin, G., \& Romeo, A.B. 1988, \aap, 195, 105

\bibitem{}  Bertin, G., Lin, C.C., Lowe, S.A., \& Thurstans, R.P. 1989a, \apj,
338, 78

\bibitem{}  \_\_\_\_\_. 1989b, \apj, 338, 104

\bibitem{} Blanchard A., Valls-Gabaud D., Mamon G., 1992, \aap, 264, 365

\bibitem{} Blitz, L., Spergel, D.N., Teuben, P.J., Hartmann, D., \& Burton,
W.B. 1998, \apj, 514, 818

\bibitem{}  Bond, J.R., Cole, S., Efstathiou, G., Kaiser, N. 1991, \apj, 379, 440

\bibitem{} Bothun, G., D., Mould, J., Schommer, R.A., \& Aaranson, M. 1985, \apj, 291, 586

\bibitem{}  Bower, R. 1991, MNRAS, 248, 332

\bibitem{}  Bressan, A., Chiosi, C., \& Fagotto, F. 1994, \apjs, 94, 63

\bibitem{}  Bruzual, G.A., \& Charlot, S. 1993, \apj, 405, 538

\bibitem{}  Burkert, A. 1995, \apj, 447, L25

\bibitem{}  Cardelli, J.A., Clayton, G.C., \& Mathis, J.S. 1989, \apj, 345, 245

\bibitem{}  Carignan, C., \& Freeman, K.C. 1985, \apj, 160, 811

\bibitem{} Casertano, S., \& van Gorkom, J.H. 1991, \aj, 101, 1231

\bibitem{}  Catelan, P., \& Theuns, T. 1996, \mnras, 282, 436

\bibitem{}  Charlot, S., Worthey, G., \& Bressan A. 1996, \apj, 457, 625

\bibitem{} Cole, S., \& Lacey, C. 1996, \mnras, 281, 716

\bibitem{}  Cole, S., Aragon-Salamanca, A., Frenk, C.S., Navarro, J., \& Zepf, S.
1994, \mnras, 271, 781

\bibitem{}  Combes, F. 1993, in ``The Formation and Evolution of Galaxies'', eds.
Mu\~{n}oz-Tu\~{n}\'{o}n \& F. S\'{a}nchez (Cambridge Univ. Press), p. 317

\bibitem{} Corsini, E.M. et al. 1998, \aap, 342, 671

\bibitem{}  Courteau, S., de Jong, R.S., \& Broeils, A.H. 1997, \apj, 457, L73

\bibitem{}  Dalcanton, J.J., Spergel, D.N., \& Summers, F.J. 1997, \apj, 482, 659

\bibitem{} de Blok, W.J.G., McGaugh, S.S. 1997, \mnras, 290, 533

\bibitem{}  de Jong, R.S. 1995, Ph.D. Thesis, University of Groningen

\bibitem{}  \_\_\_\_\_. 1996a, \aap, 313, 45

\bibitem{}  \_\_\_\_\_. 1996b, \aap, 313, 377

\bibitem{}  \_\_\_\_\_. 1996c, A\&ASS, 118, 557

\bibitem{}  de Vaucouleurs, G., et al. 1991, Third Reference Catalogue of Bright
Galaxies (Springer, Berlin Heidelberg New York)

\bibitem{}  Disney, M.J., Davies, J., \& Philipss, S. 1989, \mnras, 205, 1253

\bibitem{} Eke, V., Navarro, J.F., \& Frenk, C.S. 1998, \apj, 503, 569

\bibitem{} Fall, S.M., \& Efstathiou, G., 1980, \mnras, 193, 189

\bibitem{} Firmani, C., \& Avila-Reese, V. 2000a, \mnras, accepted (FA)

\bibitem{} Firmani, C., \& Avila-Reese, V. 2000b, preprint

\bibitem{}  Firmani, C., \& Tutukov, A.V. 1992, \aap, 264, 37

\bibitem{}  \_\_\_\_\_. 1994, \aap, 288, 713

\bibitem{}  Firmani, C., Hern\'{a}ndez, X., \& Gallagher 1996, \aap, 308, 403

\bibitem{} Flores, R.A., \& Primack, J.R. 1994, \apj, 427, L1

\bibitem{}  Flores, R.A., Primack, J.R., Blumenthal, G.R., \&\ Faber, S.M. 1993, %
\apj, 412, 443

\bibitem{} Franco, J., Santill\'an, A., \& Martos, M. 1995, in ``The
formation of the Milky Way'', G.Tenorio-Tagle, M.Prieto, \& S\'anchez, 
F. eds. (Cambridge Univ. Press), p. 515

\bibitem{} Fukugita, M., Hogan, C.J., \& Peebles, P.J.E. 1998, \apj, 
503, 518

\bibitem{} Fukushige, T., \& Makino, J. 1996, \apj, 477, L9

\bibitem{}  Gallagher, J.S., Hunter, D.A., \& Tutukov, A.V. 1984, \apj, 284, 544

\bibitem{}  Gavazzi, G. 1993, \apj, 419, 469

\bibitem{} Gavazzi, G., Pierini, D., \& Boselli, A. 1996, \aap, 312, 397

\bibitem{}  Giovanelli, R., Haynes, M.P., Herter, T.H., Vogt, N.P., da Costa,
L.N., Freudling, W., Salzer, J.J., \& Wegner, G. 1997, \aj, 113, 53

\bibitem{}  Giovanelli, R., Haynes, M., Salzer, J.J., Wegner, G., Da Costa, L.N.,
Freudling, W. 1995, \aj, 110, 1059

\bibitem{}  Giraud, E. 1986, \apj, 309, 512

\bibitem{}  \_\_\_\_\_. 1987, \aap, 174, 23

\bibitem{} Gunn, J.E. 1977, \apj, 218, 592

\bibitem{} \_\_\_\_\_. 1981, in ``Astrophysical Cosmology'', M.S. Longair, Coyne
G.V., \& H.A. Br\"{u}ck, eds. (Pontificia Academia Scientarium:Citta del
Vaticano), p.191

\bibitem{}  \_\_\_\_\_. 1987, in ``The Galaxy'', G. Gilmore \& B. Carswell, eds.
(Reidel Publishing Company), p.413

\bibitem{}  Heyl, J.S., Cole, S., Frenk, C., Navarro, J. 1995, \mnras, 274, 755

\bibitem{}  Heavens, A., \& Peacock, J. 1988, \mnras, 232, 339

\bibitem{}  Hoffman, Y. 1986, \apj, 301, 65

\bibitem{}  \_\_\_\_\_. 1988, \apj, 329,8

\bibitem{} Jimenez, R., Padoan, P., Matteucci, F., \& Heavens, A.F. 1998,
\mnras, 299, 123

\bibitem{} Jing, Y.P. 1999, \apj, submitted (astro-ph9901340)

\bibitem{} Jing, Y.P., \& Suto, Y.S. 2000, \apj, 529, L69

\bibitem{} Katz, N. 1992, \apj, 391, 502

\bibitem{}  Kauffmann, G. 1995, \mnras, 274, 161

\bibitem{}  \_\_\_\_\_. 1996, \mnras, 281, 475

\bibitem{}  Kauffmann, G., White, S.D.M., \& Guiderdoni, B. 1993, \mnras, 264, 201

\bibitem{}  Kennicutt, R.C. 1983, \apj, 272, 5

\bibitem{}  \_\_\_\_\_. 1989, \apj, 344, 6854

\bibitem{} \_\_\_\_\_. 1998, \apj, 498, 541

\bibitem{}  Kennicutt, R.C., Tamblyn, P., \& Congdon, C.W. 1994, \apj, 435, 22

\bibitem{}  Kepner, J. 1997, preprint (astro-ph/9710329)

\bibitem{}  Kodama, T., \& Arimoto, N. 1997, \aap, 320, 41

\bibitem{}  Kraan-Korteweg, R.C., Cameron, L.M., \& Tammann, G.A. 1988, \apj,
331, 620

\bibitem{}  Kudrya, Yu.N., Karachentseva, V.E., Karachentsev, I.D, \& Parnovsky,
S.L. 1997, Asrtonomy Letters, 23, 633

\bibitem{} Kull, A. 1999, \apj, 516, L5

\bibitem{}  Lacey, C.G., \& Cole S., 1993, MNRAS, 262, 627

\bibitem{}  Lacey, C.G., Guiderdoni, B., Rocca-Volmerange, \& Silk, J. 1993, %
\apj, 402, 15

\bibitem{}  Larson, R.B., \& Tinsley, B. M. 1978, \apj, 219, 46

\bibitem{}  Larson, R.B., \& Tinsley, B. M., \& Caldwell, C.N. 1980, \apj, 237,
692

\bibitem{} L\'opez-Corredoira, M., Beckman, J.E., \& Casuso, E. 1999, \aap, 351, 920

\bibitem{} Mac Low, M.-M., \& Ferrara, A. 1999, \apj, 513, 142

\bibitem{} Mac Low, M.-M., McCray, R., Norman, M.L. 1989, \apj, 337, 141 

\bibitem{} M\'arquez, I., \& Moles, M. 1999, \aap, 344, 421

\bibitem{}  McGaugh, S.S., de Blok, W.J.G. 1997, \apj, 481, 689 

\bibitem{}  Mo, H.J., Mao, S., \& White, S.D.M. 1998, \mnras, 295, 319

\bibitem{} Moore, B. 1994, Nature, 370, 629

\bibitem{} Navarro, J.F. 1998, preprint (astro-ph/9807084)

\bibitem{} Navarro, J.F., \& Steinmetz, M. 1997, \apj, 478, 13

\bibitem{} Navarro, J.F., Frenk, C.S. \& White, S.D.M. 1996, \apj, 462, 563

\bibitem{} \_\_\_\_\_. 1997, \apj, 490, 493

\bibitem{} Natarajan, P. 1999, \apj, 512, L105

\bibitem{}  Norman, C.A., Sellwood, J.A., \& Hassan, H. 1996, \apj, 462, 114

\bibitem{}  Nulsen, P.E.J., \& Fabian, A.C. 1995, \mnras, 277, 561

\bibitem{} Padmanabhan, T. 1993, ``Structure formation in the universe''
(Cambridge Univ. Press, New York)

\bibitem{}  Peletier, R., \& Balcells, M. 1996, \aj, 111, 2238

\bibitem{} Persic, M., Salucci, P., \& Stel, F. 1996, \mnras, 281, 27

\bibitem{} Pierce, M.J., \& Tully, R.B. 1992, \apj, 387, 47

\bibitem{} Portinari, L., \& Chiosi, C. 1999, \aap, accepted

\bibitem{} Prunet, S., \& Blanchard, A. 1999, \aap, accepted

\bibitem{}  Renzini, A. 1994, in ``Galaxy Formation'', J.Silk \& N. Vittorio,
eds. (Elsevier Science Publishers B.V., The Netehrlands), p.303

\bibitem{}  Roberts, M.S. 1978, \aj, 83, 1026

\bibitem{}  Roberts, M.S., \& Haynes, M.P. 1994, \araa, 32, 115

\bibitem{}  Rubin, V.C., Thonnard, N., \& Ford, W.K. 1980, \apj,
238, 471

\bibitem{}  Rubin, V.C., Burstein, D., Ford, W.K., \& Thonnard, N. 
1985, \apj, 289, 81
i
\bibitem{}  Ryden, B.S., \& Gunn, J.E. 1987, \apj, 318, 15

\bibitem{} Sancisi, R., \& van Albada, T.S. 1987, in ``Dark matter in the
universe'', IAU Symposium 117, eds. J.Kormendy, \& G.R.Knapp,
(Reidel:Dordrecht), p.67

\bibitem{} Shustov, B., Wiebe, D., \& Tutukov, A. 1997, \aap, 317, 397

\bibitem{}  Simien, F., \& de Vaucouleurs, G. 1986, \apj, 302, 564

\bibitem{} Slavin, J.D., \& Cox D.P. 1992, \apj, 392, 131

\bibitem{}  Somerville, R.S., \& Primack, J.R. 1998, preprint (astro-ph/9802268v2)

\bibitem{}  Strauss, M.A., \&\ Willick, J.A. 1995, \physrep, 261, 271

\bibitem{} Struck, C., \& Smith, D.C. 1999, \apj, accepted

\bibitem{} Sugiyama, N. 1996, ApJS, 100, 281

\bibitem{}  Tinsley, B. 1980, Fundamental Cosmic Phys., 5, 287

\bibitem{}  Toomre, A. 1964, \apj, 139, 1217

\bibitem{}  T\'{o}th, G., \& Ostriker, J.P. 1992, \apj, 389, 5

\bibitem{}  Tully, R.B. 1988, Nearby Galaxies Catalog (Cambridge, U.P.)

\bibitem{}  Tully, B., Mould, J., \& Aaranson, M. 1982, \apj, 257, 527

\bibitem{} Tully, R.B., \& Verheijen, M.A.W. 1997, \apj, 484, 145

\bibitem{} van den Bergh, S., \& Pierce, M.J. 1990, \apj, 364, 444

\bibitem{} van den Bosch, F.C. 1998, \apj, 507, 601

\bibitem{}  \_\_\_\_\_., 1999, \apj, accepted (astro-ph/9909298)

\bibitem{} Verheijen, M.A.W. 1998, PhD. Thesis, Groningen University

\bibitem{}  Vishnavatan, N. 1981, \aap, 100, L20

\bibitem{}  Wang, B., \& Heckman, T.M. 1996, \apj, 457, 645 (WH96)

\bibitem{} Weinberg, M. 1998, \mnras, 299, 499

\bibitem{} Weyl, M.L., Eke, V.R., \& Efstathiou, G. 1998, \mnras, 300, 773

\bibitem{}  White, S.D.M, \& Frenk, C.S. 1991, \apj, 379, 52

\bibitem{}  White, S.D.M., \& Rees, M.J. 1978, \mnras, 183, 341

\bibitem{}  Wyse, R. 1982, \mnras, 199, 1P

\bibitem{}  Wyse, R.F.G., Gilmore, G., \& Franx, M. 1997, \araa, 35, 637

\bibitem{} Zaroubi, S., \& Hoffman, Y. 1993, \apj, 416, 410

\bibitem{}  Zwaan, M.A., van der Hulst, J.M., de Blok, W.J/G., \& McGaugh,
S.S. 1996, \mnras, 273, 35

\end{thebibliography}
\end{document}